\documentclass[preprint]{aastex63}

\usepackage{graphicx}   
\usepackage{amsmath}
\usepackage[percent]{overpic}
\usepackage{array}      
\usepackage{verbatim}   
\usepackage{bm}

\usepackage{graphbox}

\usepackage{mwe}
\usepackage{rotating}
\usepackage{enumerate}
\usepackage{color}
\usepackage{epsfig}     
\usepackage{url}        
\usepackage{subfigure}  
\usepackage{multirow} 
\usepackage{wrapfig}
\usepackage{float}
\usepackage[hyphens]{url}
\usepackage{hyperref}
\usepackage{color}

\newcommand{\blue}{\textcolor{blue}}

\begin{document}
\title{Coronal Hole Detection and Open Magnetic Flux}

\correspondingauthor{Jon A. Linker}
\email{linkerj@predsci.com}

\author[0000-0003-1662-3328]{Jon A. Linker}
\affiliation{Predictive Science Inc., 9990 Mesa Rim Road, Suite 170, San Diego, CA 92121, USA}

\author[0000-0002-2655-2108]{Stephan G. Heinemann}
\affiliation{Max-Planck-Institut für Sonnensystemforschung, Justus-von-Liebig-Weg 3, 37077 G\"ottingen, Germany}
\affiliation{Institute of Physics, University of Graz, Universitätsplatz 5, 8010 Graz, Austria}

\author[0000-0003-4867-7558]{Manuela Temmer}
\affiliation{Institute of Physics, University of Graz, Universitätsplatz 5, 8010 Graz, Austria}

\author[0000-0003-2061-2453]{Mathew J. Owens}
\affiliation{Space and Atmospheric Electricity Group, Department of Meteorology, University of Reading, Earley Gate, P.O. Box 243, Reading RG6 6BB, UK}

\author[0000-0002-2633-4290]{Ronald M. Caplan}
\affiliation{Predictive Science Inc., 9990 Mesa Rim Road, Suite 170, San Diego, CA 92121, USA}

\author[0000-0003-1662-3328]{Charles N. Arge}
\affiliation{Heliophysics Science Division, NASA Goddard Space Flight Center, Code 671, Greenbelt, MD,
20771, USA}

\author[0000-0002-6998-7224]{Eleanna Asvestari}
\affiliation{Department of Physics, University of Helsinki, P.O. Box 64, 00014, Helsinki, Finland}

\author[0000-0001-5307-8045]{Veronique Delouille}
\affiliation{Royal Observatory of Belgium, Avenue Circulaire 3, B-1180 Bruxelles, Belgium}

\author[0000-0003-1759-4354]{Cooper Downs}
\affiliation{Predictive Science Inc., 9990 Mesa Rim Road, Suite 170, San Diego, CA 92121, USA} 

\author[0000-0001-7662-1960]{Stefan J. Hofmeister} 
\affiliation{Institute of Physics, University of Graz, Universitätsplatz 5, 8010 Graz, Austria}
\affiliation{Columbia Astrophysics Laboratory, Columbia University, 550 West 120th Street, New York, NY 10027, USA}

\author[0000-0002-0606-7172]{Immanuel C. Jebaraj}
\affiliation{Royal Observatory of Belgium, Avenue Circulaire 3, B-1180 Bruxelles, Belgium}
\affiliation{Centre for mathematical Plasma Astrophysics, Department of Mathematics, KU Leuven, Celestijnenlaan 200B, B-3001 Leuven, Belgium}

\author[0000-0001-9806-2485]{Maria S. Madjarska}
\affiliation{Max-Planck-Institut für Sonnensystemforschung, Justus-von-Liebig-Weg 3, 37077 G\"ottingen, Germany}

\date{\today}

\author[0000-0001-8247-7168]{Rui F. Pinto}
\affiliation{LDE3, DAp/AIM, CEA Saclay, 91191 Gif-sur-Yvette, France}
\affiliation{IRAP, Université de Toulouse; UPS-OMP, CNRS; 9 Av. colonel Roche, BP 44346, F-31028 Toulouse cedex 4, France}

\author[0000-0003-1175-7124]{Jens Pomoell}
\affiliation{Department of Physics, University of Helsinki, P.O. Box 64, 00014, Helsinki, Finland}

\author[0000-0002-7676-9364]{Evangelia Samara}
\affiliation{Royal Observatory of Belgium, Avenue Circulaire 3, B-1180 Bruxelles, Belgium}
\affiliation{Centre for mathematical Plasma Astrophysics, Department of Mathematics, KU Leuven, Celestijnenlaan 200B, B-3001 Leuven, Belgium}

\author[0000-0002-5681-0526]{Camilla Scolini}
\affiliation{Institute for the Study of Earth, Oceans, and Space, University of New Hampshire, Durham, NH 03824, USA}
\affiliation{University Corporation for Atmospheric Research, Boulder, CO, USA}

\author{Bojan Vr\v{s}nak}
\affiliation{Hvar Observatory, Faculty of Geodesy, University of Zagreb, Ka\v{c}i\'ceva 26, 10000, Zagreb, Croatia}

\begin{abstract}
Many scientists use coronal hole (CH) detections to infer open magnetic flux.  Detection techniques differ in the areas that they assign as  open, and may obtain different values for the open magnetic flux.  We characterize the uncertainties of these methods, by applying six different detection methods to deduce the area and open flux of a near-disk center CH observed on 9/19/2010, and applying a single method to five different EUV filtergrams for this CH.   Open flux was calculated using five different magnetic maps.  The standard deviation (interpreted as the uncertainty) in the open flux estimate for this CH $\approx$$26\%$.  However, including the variability of different magnetic data sources, this uncertainty almost doubles to $45\%$.  We use two of the methods to characterize the area and open flux for all CHs in this time period.  We find that the open flux is greatly underestimated compared to values inferred from in-situ measurements (by 2.2-4 times).   We also test our detection techniques on simulated emission images from a thermodynamic MHD model of the solar corona.  We find that the methods overestimate the area and open flux in the simulated CH, but the average error in the flux is only about $7\%$.   The full-Sun detections on the simulated corona underestimate the model open flux, but by factors well below what is needed to account for the missing flux in the observations.  Under-detection of open flux in coronal holes likely contributes to the recognized deficit in solar open flux, but is unlikely to resolve it.
\end{abstract}

\section{Introduction}
\label{sec_intro}

The solar wind is a magnetized plasma that expands outward from the solar corona to fill the interplanetary space. It plays a key role in heliophysics, providing
the medium by which solar-originating space weather-driving phenomena, such as coronal mass ejections (CMEs) and solar energetic particles, produce effects/impacts on Earth and on the surrounding space environment.  The solar wind is approximately structured into two types:  slow and fast with different sources \citep[][]{schwenn81}. Fast solar wind streams are associated with recurrent geomagnetic activity \citep{neupert_pizzo1974} and are therefore of increased research interest. They have been identified to originate from deep within coronal holes \citep{kriegeretal1973}, where the predominantly open magnetic field allows plasma to escape easily \citep{altschuleretal1972}. Along these open magnetic field lines, the  density and temperature of the outflowing plasma falls rapidly with height, causing a relatively low intensity emission of coronal holes (hereafter, CHs) in EUV and X-Ray images, or correspondingly bright in He I 10830 absorption \citep[][]{bohlin1977}. The bulk of the Sun's open magnetic flux that is measured in interplanetary space is therefore expected to originate from CH regions. However, recent investigations have shown that the open magnetic flux identified in CHs underestimates the open magnetic flux in the heliosphere deduced from in-situ measurements by a factor of two or more, referred to as the ``Open Flux Problem'' \citep{linkeretal2017,lowderetal2017,wallacetal2019}. While the fast wind is associated with the CHs themselves, the more variable slow solar wind is associated with the CH boundaries.   In theories that invoke a quasi-static origin,  the slow wind arises from regions of large expansion factor near the boundaries \citep{wang_sheeley1990,cranmeretal2007}.   Interchange reconnection  \citep[reconnection between open and closed fields,][]{crooker2002} has been suggested as the source of a dynamic slow solar wind  \citep{fisketal1998, antiochosetal2011} and would most easily occur near CH boundaries. \citet[][]{fisk20} argue that recent measurements from Parker Solar Probe \citep[PSP,][]{fox16} show that open magnetic flux is transported by interchange reconnection. Identifying and characterizing CHs and their boundaries is therefore crucial to understand the origins of the solar wind and to assess the uncertainties in the quantification of open magnetic flux.

The identification of CH regions is traditionally performed by visual inspection of image data \citep{harvey_recely2002}. In recent years, several automatic or semi-automatic routines 
have been developed for more objective results \citep{henney_harvey2005,scholl_habbal2008,krista_gallagher2009,rotteretal2012,lowderetal2014,2014_Verbeeck_SPoCA,boucheronetal2016,caplanetal2016,gartonetal2018,heinemannetal2019}. In combination with photospheric magnetic field data, the extracted CH area can be used to derive the open magnetic flux from that region. If a detection method allows a full-Sun map of CHs to be created for a given time period, then the solar open flux can be estimated entirely from observations by overlaying the CH map onto a synchronic or diachronic (often referred to as synoptic) magnetic map \citep{lowderetal2014,linkeretal2017,wallacetal2019}.  

As identified by \citet{linkeretal2017}, there are two broad categories of resolutions for this underestimate of the open flux: (1) Either the observatory magnetic maps are underestimating the magnetic flux, or (2) a significant portion of the open magnetic flux is not rooted in regions which are dark in emission. Category (1) includes possible underestimates by the magnetographs, which often disagree quantitatively \citep[e.g.,][]{rileyetal2014}, or underestimates in specific regions, such as the poorly observed polar regions \citep{rileyetal2019}. Category (2) raises the important question of how well currently available CH detection methods perform and how they compare to each other.

To address and resolve this issue, we formed an International Space Science Institute (ISSI\footnote{http://www.issibern.ch/teams/magfluxsol/}) team,  and we report the outcome of the first team meeting here. We study a well-observed low-latitude CH and its associated Carrington Rotation (CR), that occurred during solar minimum at the beginning of cycle 24 (CR2101, 2010-09-05 -- 2010-10-03). 
We investigate the uncertainties in the calculation of open magnetic flux from remote observations by exploring the variability in the results that occur when different CH detection techniques, different wavelengths, instruments, and different photospheric magnetic maps, are used. As there is no ``ground truth'' measurement for the open flux on the Sun, we use a thermodynamic MHD model \citep[e.g.,][]{mikicetal2018} to simulate the corona for this time period and produce synthetic EUV emission images.  The same analysis that was performed on the observations is repeated for the simulated data, where the ``true'' open flux is known.  The observational and model results are related to in-situ estimates of the heliospheric magnetic flux. From that we asses the overall ability of detection methods to account for solar open flux and identify potential sources of missing open flux.   

\section{Methodology and data}


\label{sec_time_period}
\begin{figure}
    \centering
    \includegraphics[width=0.9\textwidth]{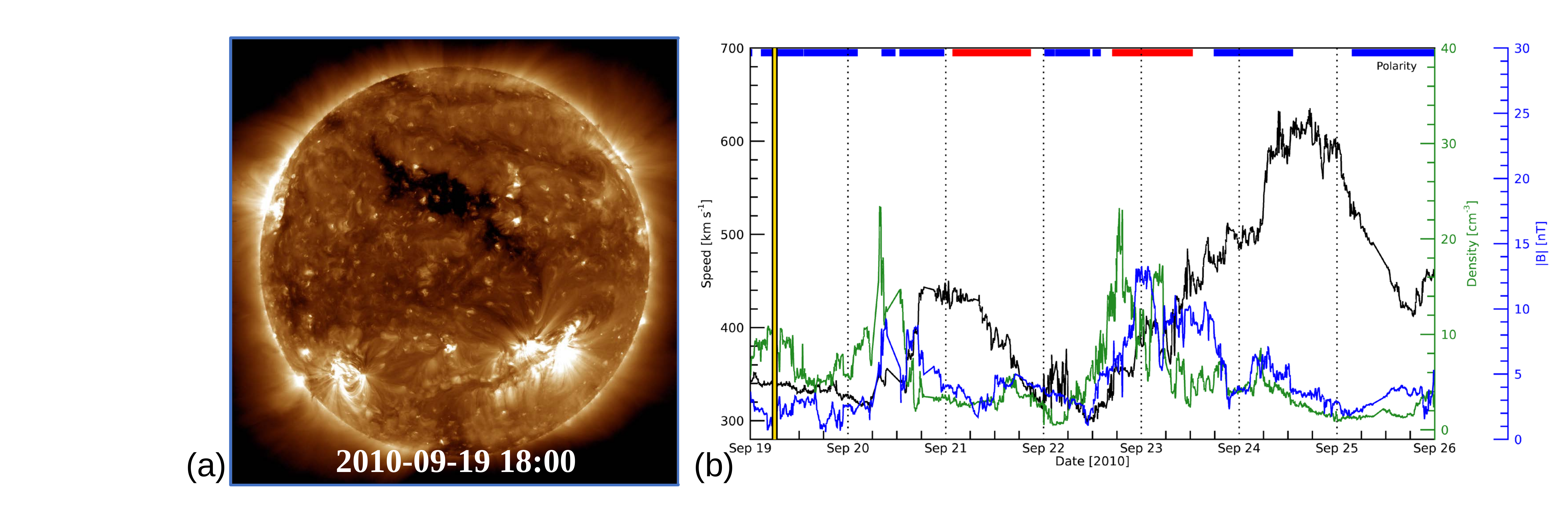}
    \caption{(a) AIA/SDO 193\AA\ full-disk image of the Sun on September 19, 2010.  The CH selected for this study is easily visible in the northern hemisphere. (b) In-situ signatures of the associated solar wind (data provided by the OMNI database), where the black line shows the solar wind bulk velocity, the green line represents the plasma density and the blue line is the magnetic field strength. The red-blue colored bars on the top represent the polarity of the in-situ measured magnetic field calculated after \cite{neugebauer02}, with red being positive and blue negative polarity, and the time of the SDO observation corresponds to the yellow vertical line.}
    \label{fig:euvimg}
\end{figure}

We focus on one particular CH observed on September 19, 2010, and the surrounding time period (CR2101, 2010-09-05 -- 2010-10-03). 
We selected this time period and CH based on the following criteria i) availability of high-resolution SDO/AIA and low noise SDO/HMI data, ii) a solar minimum time period, when there is less solar activity and the coronal configuration is simpler, iii) an isolated CH, i.e., not connected to a polar coronal hole nor surrounded by strong active regions, with comparatively well-defined boundaries at the solar surface for keeping projection effects to a lower limit, iv) clear signatures of the associated high-speed stream from in-situ data. 
Figure~\ref{fig:euvimg} shows the Sun on September 19, 2010 with the CH under study located in the central part of the solar disk (panel a) as well as the related solar wind high-speed stream at 1 AU from in-situ data (panel b). 

\subsection{CH Detection Methods}
\label{sec_detect_method}

There are now several automated and semi-automated methods for detecting CH boundaries from emission images, and these are often used to identify regions of open magnetic flux. At the present time, the accuracy of these methods is unclear, and there has been little intercomparison between the methods.
The uncertainty of these methods is therefore an open question that directly impacts the larger question of why coronal estimates of open flux disagree with inferences from in-situ measurements.

\begin{figure}
    \centering
        \includegraphics[width=0.65\textwidth]{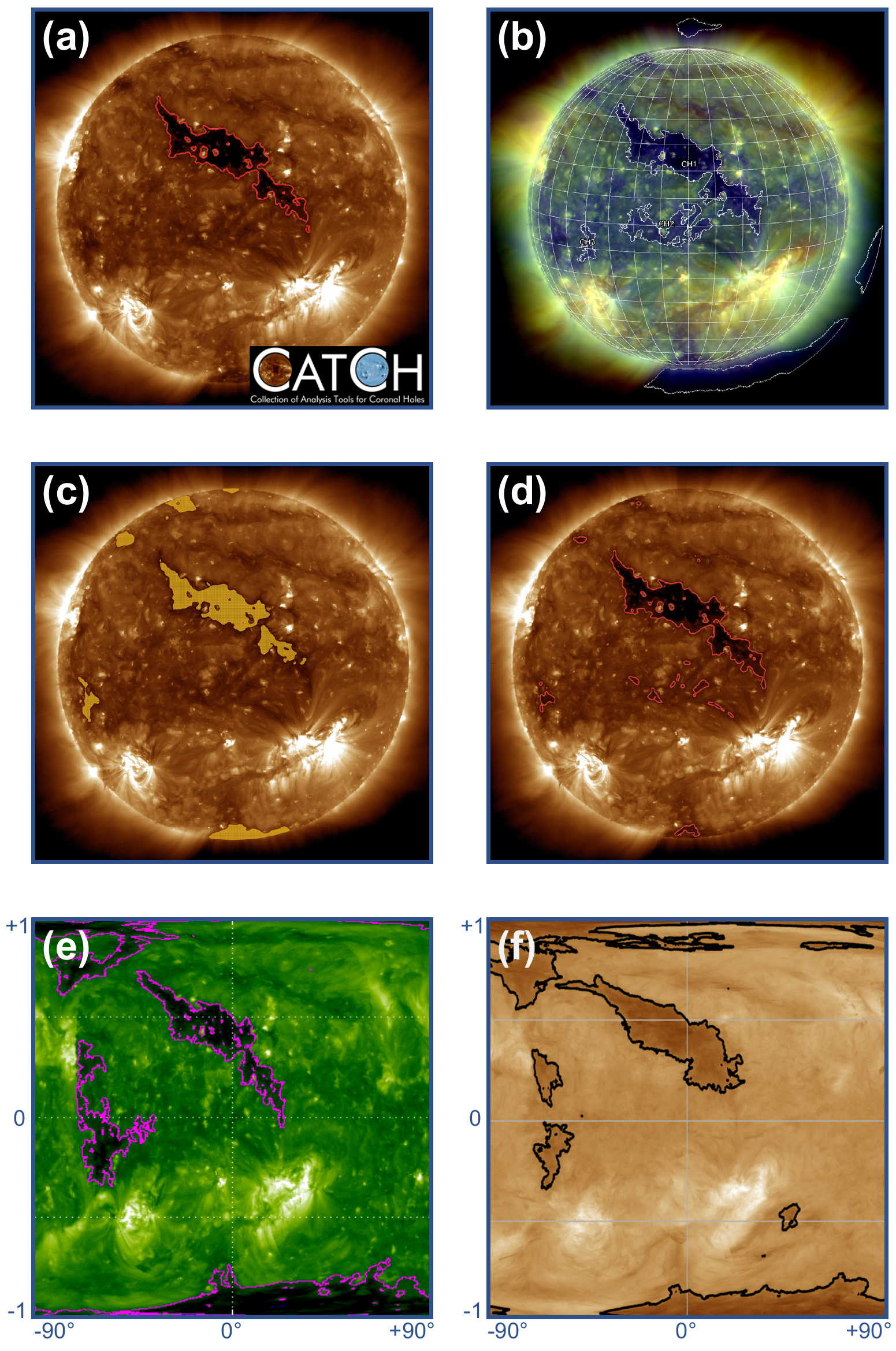}
    \caption{Examples of the CH detection and extraction methods used in this study applied on EUV image data. Images are for the CH of interest that crossed the central meridian on September 19, 2010. Panels (a) through (f) feature CATCH, CHIMERA, SPoCa, THR, PSI-SYNCH and PSI-MIDM respectively.  To focus on the CH of interest, full-sun PSI-SYNCH and PSI-MIDM detection maps are cut out in a $\pm 90^{\circ}$ longitude vs. sine-latitude Stonyhurst heliographic projection.}
    \label{fig:overview}
\end{figure}

We apply six different but commonly used methods to this CH, and estimate the uncertainties in CH detection, which in turn leads to uncertainties in the observed open flux. The extraction methods used are: simple thresholding \citep[THR;][]{rotteretal2012, krista_gallagher2009}, the Spatial Possibilistic Clustering Algorithm \citep[SPoCA,][]{2014_Verbeeck_SPoCA}, Synchronic Coronal Hole Mapping \citep[PSI-SYNCH,][]{caplanetal2016}, the Minimum Intensity Disk Merge \citep[PSI-MIDM,][]{caplanetal2019}, the Coronal Hole Identification via Multi-thermal Emission Recognition Algorithm  \citep[CHIMERA,][]{gartonetal2018}, and the Collection of Analysis Tools for Coronal Holes \citep[CATCH,][]{heinemannetal2019}. The CH area extraction methods are applied on high-resolution EUV data in several wavelengths (171\AA, 193\AA, 211\AA) from the Atmospheric Imaging Assembly aboard the Solar Dynamics Observatory \citep[AIA/SDO;][]{lemen12}. The 193\AA~wavelength range is particularly favorable for the detection of CHs due to the strong contrast between the low intensity CH region and brighter surrounding quiet corona.  In addition to AIA, we also use SWAP/PROBA2 174\AA\ \citep{seaton13}, XRT/HINODE \citep{golub07}, and 195\AA\ data from the EUVI instrument \citep[][]{wuelser04} aboard the STEREO spacecraft \citep[][]{kaiser08}. 
The different algorithms and methods are briefly described in the following. Examples applying the different extraction methods are shown in Figure~\ref{fig:overview}.

\subsubsection{Simple Thresholding (THR)}
\citet{rotteretal2012} present a CH extraction method that applies simple intensity thresholding, based on work described in \cite{vrsnak07} and \cite{krista_gallagher2009}. Following that approach, we use a threshold of 35\% of the median solar disk intensity to extract dark coronal features from SDO/AIA 193\AA~images. The $35\%$ median intensity threshold has been found to give consistent and reasonable results for CH boundaries, especially near the maximum of solar cycle 24 \citep{hofmeisteretal2017,heinemannetal2019}. 

\subsubsection{SPoCA}

The SPoCA-CH-suite \citep{2014_Verbeeck_SPoCA} is a set of segmentation procedures that allows decomposition of an EUV image into regions of similar intensity, typically active regions, CHs, and quiet Sun. It relies on an iterative clustering algorithm called fuzzy C-means, which minimizes the variance in each cluster. Typically, the CH class corresponds to the class whose center has the lowest pixel intensity value. The SDO Event Detection System runs the SPoCA-suite to extract CH information from AIA images in the 193\AA~passband, and uploads the entries every four hours to the Heliophysics Events Knowledgebase~\citep{2012SoPh..275...67H}. The code for the SPoCA-suite is available at \url{https://github.com/bmampaey/SPoCA}.
  
\subsubsection{PSI-SYNCH \& PSI-SYNOPTIC}
The goal of the PSI-SYNCH algorithm \citep{caplanetal2016} is to create, as accurately as possible, synchronic EUV and CH maps for the entire Sun. It was originally developed and applied to the 2010-2014 time period when most or all of the Sun was visible in EUV from the NASA STEREO and SDO spacecraft.  All of the PSI methods emphasize pre-processing of the image data.  PSI-SYNCH applies a point spread function (PSF) deconvolution to the full-disk images to remove stray light, especially in the CHs, and produces non-linear limb brightening correction factors and inter-instrument transformation factors using a running one year average of disk data. The CH detection is applied to each disk image using a dual-threshold region growing algorithm called EZSEG (image segmentation code). After the detection, the results for each disk are merged together into a single synchronic full-Sun CH map. Results, as well as the open-source EUV pre-processing and EZSEG codes are made available at \url{http://www.predsci.com/chd}. To produce EUV and CH maps that tend to be more continuous for dark structures, each disk image is first mapped to its own Carrington map, and then the three maps are merged by taking the lowest intensity values of the overlap.  For detection on the CH of interest, the maps are cut out in a $\pm 90^{\circ}$ longitude vs. sine-latitude Stonyhurst heliographic projection.

PSI-SYNCH provides full-Sun synchronic maps for those time periods when combined SDO and STEREO images cover the entire Sun's surface (2011-2014).
In September 2010, a portion of the Sun's surface on the backside of the Sun was not visible from the STEREO spacecraft. Therefore, combined SDO and STEREO images cannot be used to generate a full-Sun map.  To provide a full-Sun map for CR2101 from which we can estimate the Sun's total open flux, we use PSI-SYNOPTIC.   PSI-SYNOPTIC uses the same detection methodology, but instead combines images over an entire solar rotation, with pixels weighted in longitude by their proximity to disk center at the time of the observation.  This diachronic map uses only SDO AIA observations.

\subsubsection{PSI-MIDM}
A challenge for CH detection and area extraction based on single images stems from the fact that EUV line-of-sight observations flatten the three-dimensional structure in the low corona, which can cause nearby bright structures to obstruct CHs.  To mitigate this obstruction, an alternative method to combine EUV disk images into a full-Sun map, referred to as the Minimum Intensity Disk Merge (MIDM), is used.  This builds on the PSI-SYNCH and PSI-SYNOPTIC approaches, using an arbitrary number of vantage points and/or images over time.  PSI-MIDM  \citep{caplanetal2019} uses the same EZSEG algorithm as PSI-SYNCH for detection.
Instead of using centrally weighted latitude strips as in PSI-SYNOPTIC, MIDM takes full-disk images in the image- or time-sequence and merges them based on which pixels in the overlap exhibit the minimum intensity.  This allows any part of the CH area observed from any vantage point or time in the image sequence to be seen in the final map.  This can be performed even if only a single vantage point is available, by combining images taken over time  (i.e. SDO AIA observations). This creates a trade-off between detecting rapid CH evolution versus revealing portions of CHs obscured by bright loops (as can occur near active regions). For this study, we created a PSI-MIDM map and detection using SDO AIA observations at a 6 hour cadence over all of CR 2101.  As with PSI-SYNCH, a $\pm 90^{\circ}$ longitude vs. sine-latitude Stonyhurst heliographic projection is used to focus on the selected CH.


\subsubsection{CHIMERA}
CHIMERA \citep{gartonetal2018} uses the three SDO/AIA passbands where CHs are predominantly visible (171\AA, 193\AA, and 211\AA) to segment dark structures. The extraction is based on the ratios and magnitudes of the emission from each passband.  CHIMERA is an automated CH detection and extraction algorithm and derives robust boundaries which are continuously presented at \url{solarmonitor.org}.

\subsubsection{CATCH}
The recently developed CATCH algorithm \citep{heinemannetal2019} is a threshold-based CH extraction method, which uses the intensity gradient along the CH boundary to modulate the extraction threshold. By minimizing the change in the extracted area between similar thresholds, a stable boundary can be found. CATCH also provides uncertainty estimations for all parameters. Due to its concept and set-up, CATCH can be applied to any intensity-based EUV filtergram to extract low intensity regions on the solar surface. The tool is publicly available at \url{https://github.com/sgheinemann/CATCH}, including a link to the VizieR catalogue - a sample of more than 700 CHs, ready for statistical analysis.

\subsection{Open flux derivation at the Sun}\label{sec_openflux}
Photospheric magnetic maps show significant variability, from both the underlying measurements at different observatories and the method of map preparation. To address and quantify this issue, we use five different magnetic map products to calculate the open flux within the extracted CH boundaries. We obtain an estimate of the open magnetic flux for each CH detection by overlaying the extracted CH boundaries on a photospheric magnetic map taken at approximately the same time as the emission images. We integrate the signed magnetic flux in each boundary (and also obtain the signed average magnetic field) for each detection, using synoptic maps of the photospheric magnetic field from three different observatories - Michelson Doppler Imager \citep[MDI;][]{scherrer95} aboard the Solar and Heliospheric Observatory \citep[SOHO;][]{domingo95}, Helioseismic and Magnetic Imager \citep[HMI;][]{schou12} aboard SDO (720s-HMI), and the ground-based Global Oscillation Network Group instruments  \citep[GONG;][]{harvey96}. As these are derived from line-of-sight (LOS) magnetograms, the radial magnetic field ($B_r$) is obtained under the frequently used assumption that the field is radial where it is measured in the photosphere \citep{wang_sheeley1992}. Additionally, we used magnetic maps generated with the Air Force Data Assimilative Photospheric flux Transport (ADAPT) model \citep{argeetal2010,hickmannetal2015} using both HMI and GONG full-disk magnetograms as input, for a total of five different magnetic flux inputs. We note that the ADAPT model multiplied the HMI values by 1.35, and the GONG values by 1.85, prior to assimilation. All of the magnetic data were formatted to the same projection and resolution as the detected CH boundaries.

\subsection{Derivation of the Heliospheric Magnetic Field}
\label{sec_derive_helio}
Spacecraft with in-situ instruments directly measure the heliospheric magnetic field (HMF) at a single point in space. The unsigned magnetic flux threading a heliocentric sphere with measurement radius, $r$, can therefore be estimated by $\Phi_{r} = 4 \pi r^2 |B_R|$, where $B_R$ is the radial component of the HMF, if it is assumed that the single-point measurement of $|B_R|$ is representative of all latitudes and longitudes. From near-Earth space, longitudinal structure of $|B_R|$ can be measured by considering an entire Carrington rotation period and assuming the corona and HMF do not evolve significantly over this time interval. Latitudinal invariance in $|B_R|$ (scaled for heliocentric distance) was confirmed by the $Ulysses$ spacecraft on all three of its orbits \citep[][]{lockwood04,smith08}. Thus the assumption that single-point measurements of the HMF can be used to estimate $\Phi_{r}$ appears to be valid. This has been demonstrated empirically by \citet[][]{owens08}. However, there is an additional issue that $\Phi_{1AU}$ may not be equal to the unsigned flux threading the solar source surface, $\Phi_{SS}$, which is the typical definition of open solar flux (OSF). If the HMF becomes folded or inverted within the heliosphere, $\Phi_{1AU} > \Phi_{SS}$. As suprathermal electrons move anti-Sunward on a global scale \citep[][]{pilipp87}, sunward motion can be used to identify times when the HMF is locally folded \citep[][]{crooker04}. For calculating the heliospheric magnetic field, we take in-situ plasma and magnetic field measurements from the Advanced Composition Explorer \citep[ACE;][]{stone98} and its Solar Wind Ion Composition Spectrometer \citep[SWICS;][]{gloeckler98}, Solar Wind Electron Proton Alpha Monitor \citep[SWEPAM;][]{mccomas98} and Magnetometer Experiment \citep[MAG;][]{smith98}. Two of the CH detection methods (PSI-MIDM and PSI-SYNCH/SYNOPTIC) provide CH detection and open flux estimates over the entire Sun's surface. We employ these methods to estimate the amount of open flux in CR2101, and compare these results with estimates of the interplanetary magnetic flux for this time period. 

\subsection{Simulating the open flux at the Sun}
While remote solar observations allow us to infer the open solar magnetic flux in the solar corona, we are not able to measure it directly. In an alternative approach, we employ a thermodynamic MHD model \citep[e.g.,][]{mikicetal2018} for CR2101, and we create a sequence of simulated AIA images for the same view point of the real spacecraft over the course of the rotation.  We apply the CH detection methods to the simulated data, and compare with the ``true'' open flux (known from the model) to further assess the effectiveness of detection methods in accounting for open flux. The thermodynamic MHD model is briefely described in the Appendix.

\section{Inferring Open Flux from Coronal and Heliospheric Observations}
\label{sec_detect_results}
\subsection{Coronal Hole Detection}
\begin{figure}
    \centering
    \includegraphics[width=\textwidth]{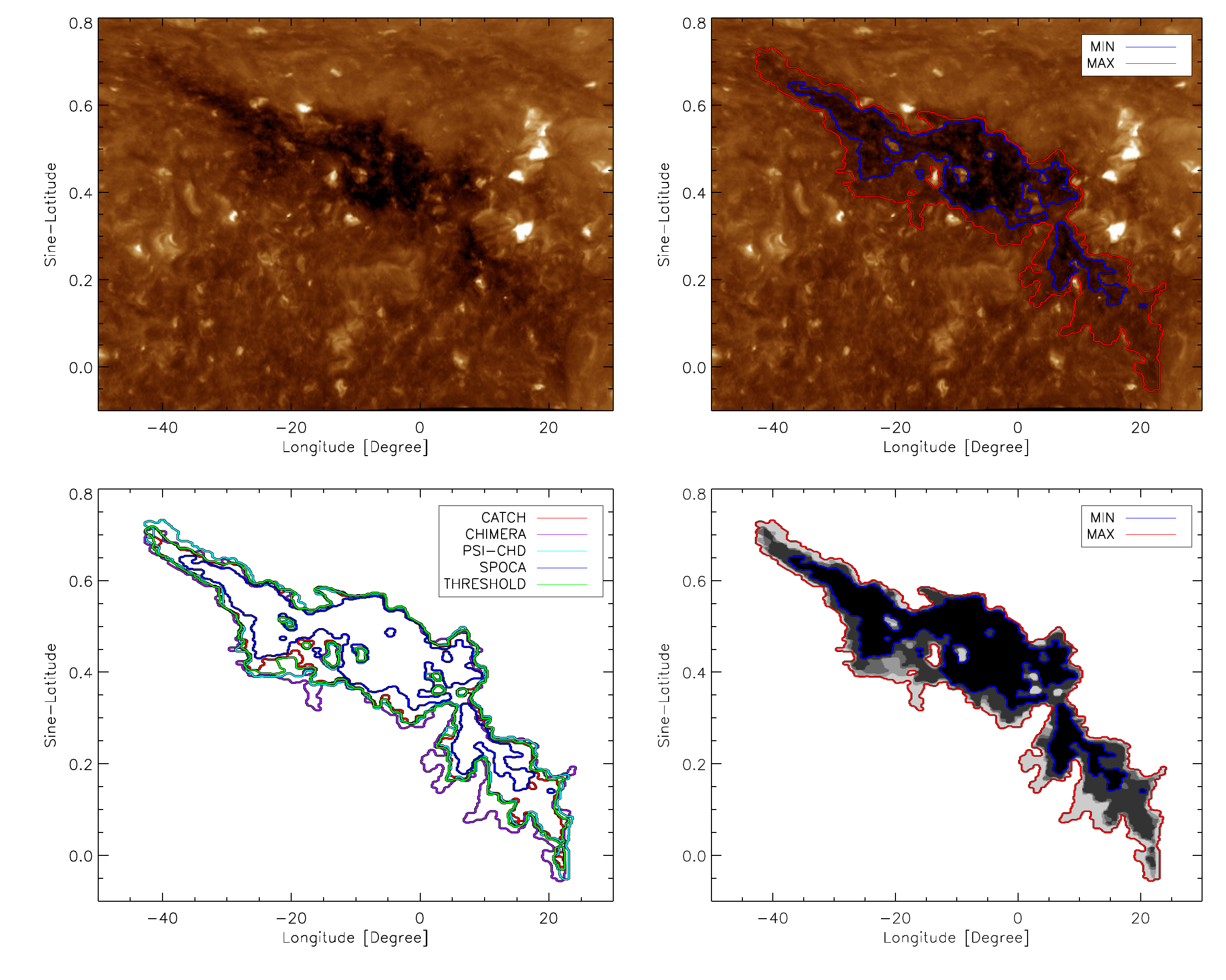}
    \caption{Detection of CH boundaries using different methods: (a) The AIA/SDO $193$\AA\ filtergram. (b) The AIA/SDO $193$\AA\ filtergram overlaid with the minimum and maximum boundary constructed by overlaying the 6 individual boundaries. (c) The extracted CH boundary from CATCH using AIA/SDO 193\AA; CHIMERA using AIA/SDO 171/193/211\AA; PSI-SYNCH using AIA/SDO 193\AA\ and EUVI/STEREO 195\AA; SPoCA using AIA/SDO 193\AA; thresholding (with $35\%$ of the median solar disk intensity); and PSI-MIDM using multiple AIA/SDO 193\AA\ images. (d) Stacked map of the 6 binary extracted CHs.   Darker pixels represent more agreement amongst the detections. }

    \label{fig:methods}
\end{figure}

We performed six detections for the selected CH using CATCH, CHIMERA, PSI-SYNCH, SPoCA, THR and PSI-MDIM. The detections were each done using their native input data and projection. For inter-comparison, the extracted CH boundaries were projected to Carrington longitude at $1^{\circ}$ per pixel and heliographic sine-latitude at $\tfrac{1}{90}$ per pixel, and smoothed using spherical morphological operators of size 3 \citep{heinemannetal2019}.  The resulting maps contained equal area pixels of roughly $9.4 \times 10^{7}$\,km$^{2}$. The comparison between the extracted boundaries is shown in Figure~\ref{fig:methods}. We find that the average extracted area is $8.89\pm 2.35 \times 10^{10}$\,km$^{2}$, with the SPoCA extraction providing the smallest value ($4.90 \times 10^{10}$\,km$^{2}$), and the largest value ($11.94\times 10^{10}$\,km$^{2}$) obtained from the PSI-MIDM extraction. The areas of the maximum and minimum extractions differ by a factor of $>2$ and the uncertainty, estimated as the standard deviation of the mean of all CH areas, is roughly $\sigma_{\rm A,d}\approx26\%$.\\

\begin{figure}
    \centering
        \includegraphics[width=\textwidth]{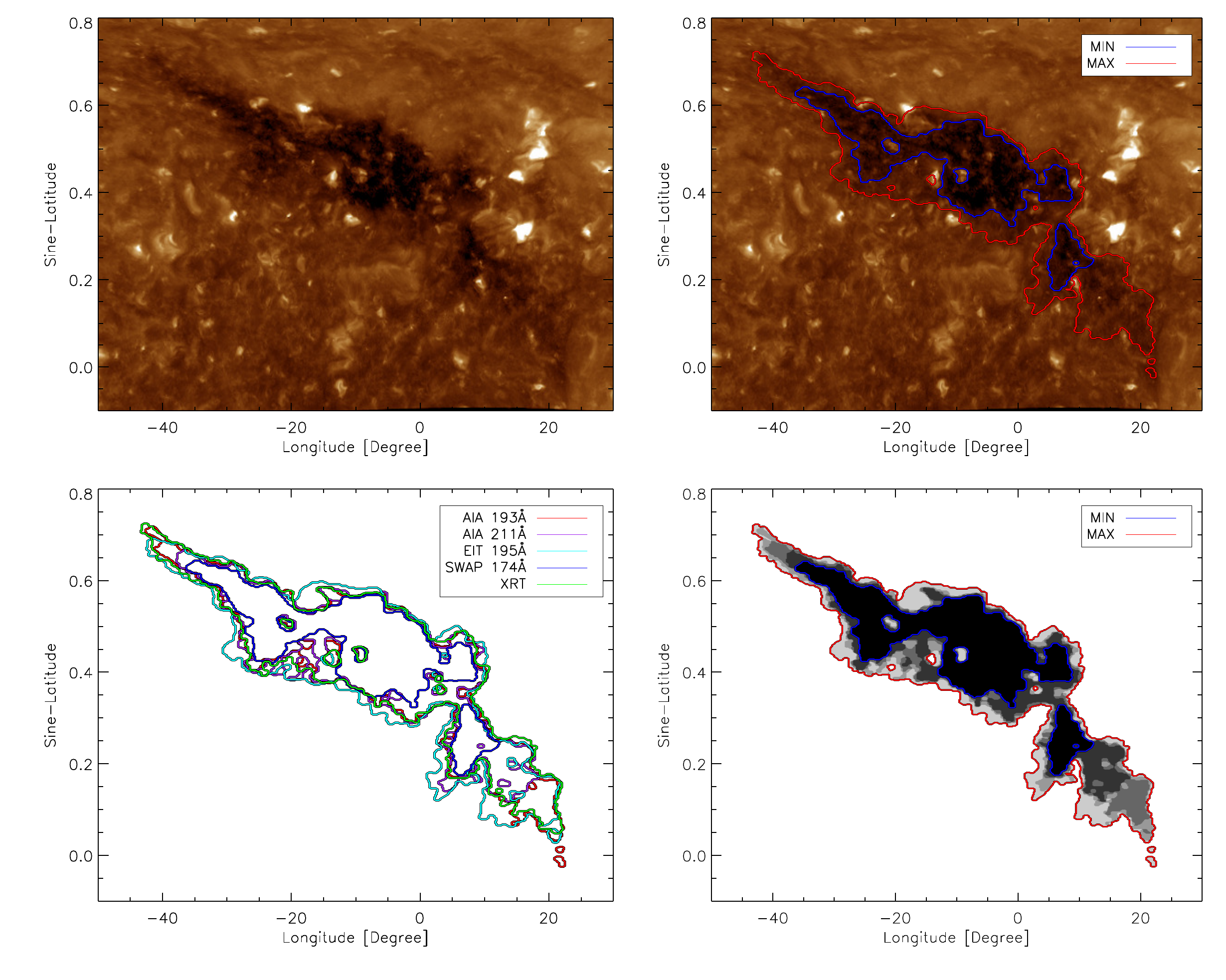}
    \caption{CH boundaries extracted using CATCH on different input EUV filtergrams. The panels show: (a) The AIA/SDO $193$\AA\ filtergram. (b) The AIA/SDO $193$\AA\ filtergram overlaid with the minimum and maximum boundaries constructed by overlaying the 5 individual boundaries. (c) The different boundaries extracted by the means of CATCH using input data from AIA/SDO 193\AA\ and 211\AA , EIT/SOHO 195\AA , SWAP/PROBA2 174\AA\ and XRT/HINODE. (d) Stacked map of the 5 binary extractions (where darker pixels represent an agreement of more extractions).}
    \label{fig:catch}
\end{figure}

To further explore the uncertainties in CH detection, we investigated how the extracted CH boundary varied for different wavelengths and instruments.  To accomplish this task, we used CATCH, which employs a detection methodology that performed close to the mean value of all the methods (cf.\,Table~\ref{tab:methods}) and moreover, is easily applicable to all intensity-based images.  We applied CATCH to five sets of input data
(193\AA\ AIA/SDO, 211\AA\ AIA/SDO, 195\AA\ EIT/SOHO, 174\AA\ SWAP/PROBA2 and XRT/HINODE). The comparison between the extracted boundaries is shown in Figure~\ref{fig:catch}.  We find the average detected CH area to be $7.83 \pm 2.00 \times 10^{10}$\,km$^{2}$ for the five different data sets.  The values range from $10.23 \times 10^{10}$\,km$^{2}$ (from EIT 195\AA\ data) to $4.77 \times 10^{10}$\,km$^{2}$ (from 174\AA\ SWAP data) which is about a factor of two. The uncertainty here is roughly in the same order as when using different extraction methods of about $\sigma_{\rm A,\textsc{catch}}\approx26\%$ of the mean. The low value obtained with the 174\AA\ wavelength should be viewed with some caution.  This line forms at a lower temperature range and at lower coronal heights than the other lines, and therefore may image different structures in the CH.

\begin{figure}
    \centering
     \includegraphics[width=\textwidth]{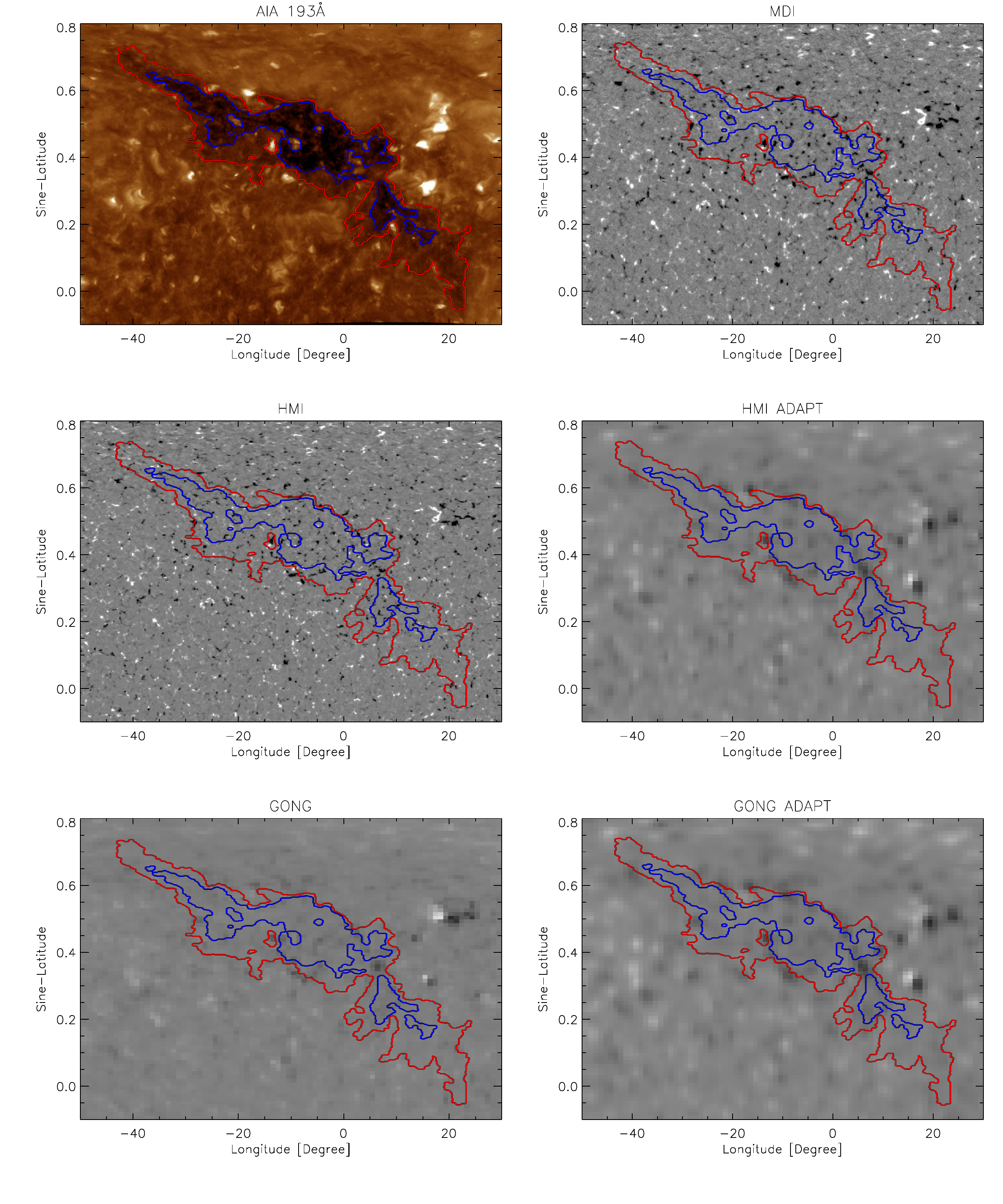}
    \caption{The CH under study (a) shown in EUV and (b--f) in the different photospheric magnetograms. The minimal and maximum CH boundaries are overlaid as blue and red contours, respectively (cf.\,Figure~\ref{fig:methods}). The magnetograms are scaled to $\pm50$\,G.}
    \label{fig:mag}
\end{figure}

To calculate the open flux within the extracted CH boundaries, we employ five different photospheric magnetic maps (section~\ref{sec_openflux}). Figure~\ref{fig:mag} shows the five different magnetic maps (MDI, HMI, GONG, GONG-ADAPT, HMI-ADAPT) overlaid with the minimum and maximum stacked CH boundaries.  

When applied to an individual magnetic map, we find that the variations in the signed mean magnetic field density are rather small ($\sigma_{\rm{B}i} < 9\%$) between different extracted boundaries (varying both the detection method and the input data).  However, large deviations are derived between the different magnetic maps with a mean magnetic field density of $-2.78 \pm 0.67$\,G within a range of $-2.0$ to $-3.5$\,G. This is equivalent to an uncertainty of $\sigma_{\rm B}\approx24\%$. When calculating the open flux from the CH area, $A$, and underlying magnetic field, $B$, as $\Phi = A \times B $ for each extraction and each map, we find an average of $(-23.59 \pm 10.75) \times 10^{20}$\,Mx ($\sigma_{\Phi,d} \approx 46\%$) for the different CH extraction methods and $(-22.35 \pm 9.53) \times 10^{20}$\,Mx ($\sigma_{\Phi,\textsc{catch}} \approx 43\%$) for the different input data with one extraction method. The values found range from $-10.9 \times 10^{20}$\,Mx to $-35.32 \times 10^{20}$\,Mx between all CH extractions. 

Table~\ref{tab:methods} summarizes the CH properties using different extraction methods, magnetic field maps and native input data. Table~\ref{tab:catch} lists the CH properties using CATCH with different input data and magnetic field maps.  We see that the ADAPT maps (using either GONG or HMI magnetograms) generally provide the highest estimates of the mean magnetic field density and magnetic flux. This is likely due to the multiplication factors applied to the input magnetograms during assimilation, which is performed in part to counter perceived underestimates of the magnetic flux.
\begin{table}
\centering
\caption{CH parameters derived from different extraction methods (CATCH, CHIMERA, SPoCA, PSI-SYNCH, PSI-MIDM and THR): area, $A$, intensity in 193\AA~image data, $I_{\rm 193}$, location of the center of mass, CoM (longitude and latitude), mean magnetic field strength, $B$, and magnetic flux, $\Phi$, for each magnetic map (MDI, HMI, HMI ADAPT, GONG, GONG ADAPT).}\label{tab:methods}
\begin{tabular}{l | c c c c c c| c } 
 & CATCH & CHIMERA & SPoCA & PSI-SYNCH & PSI-MIDM & THR & Mean  \\ \hline
A [$10^{10}$ km$^{2}$]& $8.2$& $10.2$& $4.9$ & $9.3$ & $11.9$ &$8.8$ & $ 8.9\pm2.3$  \\
I$_{193}$ [DN] & $52.2$& $58.7$& $46.4$ & $53.8$ & $65.8$ &$53.0$ & $ 55.0\pm6.6$  \\
CoM$_{\mathrm{\textsc{lon}}}$ [$^{\circ}$] & $-7.0$& $-6.3$& $-9.0$ & $-7.5$ & $-7.8$ &$-6.9$ & $ -7.4\pm0.9$  \\
CoM$_{\mathrm{\textsc{lat}}}$ [$^{\circ}$] & $23.3$& $21.6$& $25.6$ & $23.2$ & $24.5$ &$23.0$ & $ 23.5\pm1.4$  \\ \hline
B$_{\mathrm{\textsc{mdi}}}$ [G]& $-2.8$& $-2.7$& $-2.9$ & $-2.6$ & $-2.7$ &$-2.7$ & $ -2.7\pm0.1$  \\
B$_{\mathrm{\textsc{hmi}}}$ [G]& $-2.1$& $-1.9$& $-2.2$ & $-2.0$ & $-1.9$ &$-2.1$ & $ -2.0\pm0.1$  \\
B$_{\mathrm{\textsc{hmi~adapt}}}$ [G]& $-3.1$& $-2.8$& $-3.3$ & $-3.1$ & $-2.8$ &$-3.1$ & $ -3.0\pm0.2$  \\
B$_{\mathrm{\textsc{gong}}}$ [G]& $-2.4$& $-2.2$& $-2.5$ & $-2.3$ & $-2.2$ &$-2.3$ & $ -2.3\pm0.1$  \\
B$_{\mathrm{\textsc{gong~adapt}}}$ [G]& $-3.5$& $-3.0$& $-3.7$ & $-3.3$ & $-3.0$ &$-3.4$ & $ -3.3\pm0.3$  \\ \hline
$\Phi_{\mathrm{\textsc{mdi}}}$ [$10^{20}$ Mx]& $-22.5$& $-27.7$& $-14.0$ & $-24.7$ & $-32.5$ &$-23.8$ & $ -24.2\pm6.1$  \\
$\Phi_{\mathrm{\textsc{hmi}}}$ [$10^{20}$ Mx]& $-17.4$& $-19.7$& $-10.9$ & $-19.1$ & $-22.6$ &$-18.2$ & $ -18.0\pm3.9$  \\
$\Phi_{\mathrm{\textsc{hmi~adapt}}}$ [$10^{20}$ Mx]& $-25.7$& $-28.9$& $-16.2$ & $-28.6$ & $-33.3$ &$-27.1$ & $ -26.6\pm5.7$  \\
$\Phi_{\mathrm{\textsc{gong}}}$ [$10^{20}$ Mx]& $-19.6$& $-22.2$& $-12.4$ & $-21.3$ & $-26.0$ &$-20.6$ & $ -20.4\pm4.5$  \\
$\Phi_{\mathrm{\textsc{gong~adapt}}}$ [$10^{20}$ Mx]& $-28.2$& $-30.9$& $-17.9$ & $-30.6$ & $-35.3$ &$-29.7$ & $ -28.8\pm5.8$  \\ \hline

\hline
\end{tabular}

\end{table}

\begin{table}
\centering
\caption{CH parameters derived from CATCH extraction and using different EUV filtergrams: area, $A$, intensity in image data, $I_{\rm 193}$, center of mass, CoM (latitude and longitude), mean magnetic field strength, $B$, and magnetic flux, $\Phi$, for each magnetic map (MDI, HMI, HMI ADAPT, GONG, GONG ADAPT). }\label{tab:catch}
\begin{tabular}{l | c c c c c | c } 
 & AIA 193\AA\ & AIA 211\AA\ & EIT 195\AA\ & SWAP 174\AA\ & XRT & Mean  \\ \hline
A [$10^{10}$ km$^{2}$]& $8.2$& $7.4$& $10.2$ & $4.8$ & $8.6$ &$7.8\pm2.0$  \\
I$_{193}$ [DN] & $52.2$& $51.8$& $58.5$ & $47.2$ & $52.4$ &$52.4\pm4.0$  \\
CoM$_{\mathrm{\textsc{lon}}}$ [$^{\circ}$] & $-7.0$& $-8.1$& $-8.0$ & $-10.4$ & $-7.7$ &$-8.3\pm1.3$  \\
CoM$_{\mathrm{\textsc{lat}}}$ [$^{\circ}$] & $23.3$& $24.2$& $23.2$ & $26.3$ & $23.8$ &$24.2\pm1.3$  \\ \hline
B$_{\mathrm{\textsc{mdi}}}$ [G]& $-2.8$& $-2.6$& $-3.1$ & $-3.0$ & $-2.9$ &$-2.9\pm0.2$  \\
B$_{\mathrm{\textsc{hmi}}}$ [G]& $-2.1$& $-1.9$& $-2.3$ & $-2.4$ & $-2.3$ &$-2.2\pm0.2$  \\
B$_{\mathrm{\textsc{hmi~adapt}}}$ [G]& $-3.1$& $-3.2$& $-3.1$ & $-3.5$ & $-3.2$ &$-3.2\pm0.2$  \\
B$_{\mathrm{\textsc{gong}}}$ [G]& $-2.4$& $-2.4$& $-2.4$ & $-2.8$ & $-2.5$ &$-2.5\pm0.2$  \\
B$_{\mathrm{\textsc{gong~adapt}}}$ [G]& $-3.5$& $-3.6$& $-3.4$ & $-3.9$ & $-3.5$ &$-3.6\pm0.2$  \\ \hline
$\Phi_{\mathrm{\textsc{mdi}}}$ [$10^{20}$ Mx]& $-22.5$& $-19.4$& $-31.4$ & $-14.2$ & $-25.2$ &$-22.5\pm6.4$  \\
$\Phi_{\mathrm{\textsc{hmi}}}$ [$10^{20}$ Mx]& $-17.4$& $-14.3$& $-23.4$ & $-11.5$ & $-19.6$ &$-17.2\pm4.6$  \\
$\Phi_{\mathrm{\textsc{hmi~adapt}}}$ [$10^{20}$ Mx]& $-25.7$& $-23.6$& $-31.9$ & $-16.8$ & $-27.9$ &$-25.2\pm5.6$  \\
$\Phi_{\mathrm{\textsc{gong}}}$ [$10^{20}$ Mx]& $-19.6$& $-18.0$& $-24.4$ & $-13.2$ & $-21.2$ &$-19.3\pm4.1$  \\
$\Phi_{\mathrm{\textsc{gong~adapt}}}$ [$10^{20}$ Mx]& $-28.2$& $-26.4$& $-34.4$ & $-18.4$ & $-30.0$ &$-27.5\pm5.9$  \\ \hline

\hline
\end{tabular}
\end{table}

The uncertainty in the open flux from a CH can, in principle, be divided into the uncertainty from the CH extraction ($\sigma_{\Phi i} \approx 26\%$; cf.\,Table~\ref{tab:methods} \& Table~\ref{tab:catch}) and the differences/uncertainties between different magnetograms. From this we can conclude that for 
a typical extraction
method the uncertainty in the open flux derivation on a well-observed CH is $\sigma_{\rm \Phi}\approx 43-46\%$. 
\subsection{Comparison with heliospheric open flux}
\begin{figure}
    \centering
     \includegraphics[width=\textwidth]{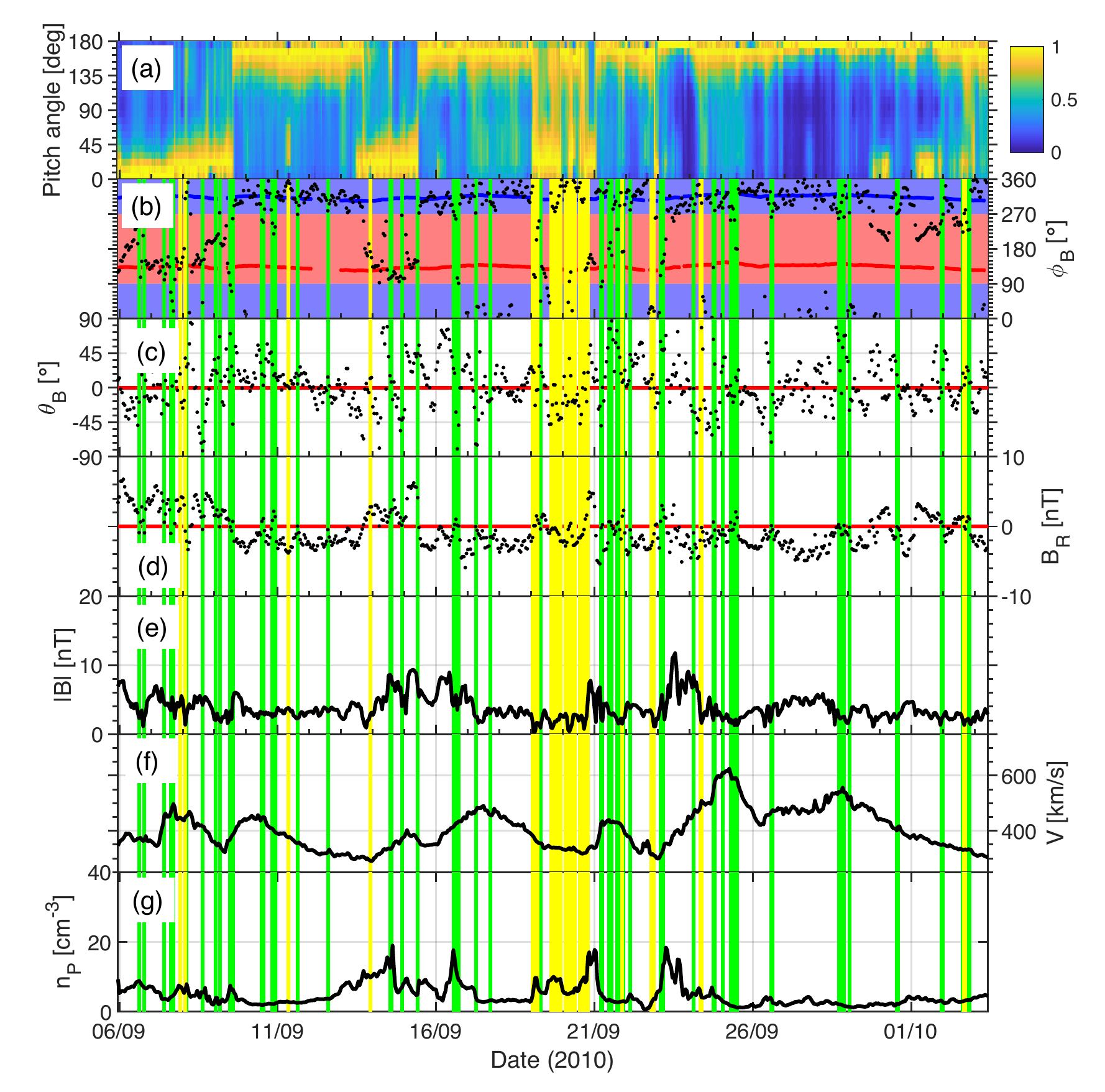}
    \caption{In-situ observations during CR2101 from the ACE spacecraft. (a) Pitch angle distribution of 272 eV electrons, with the intensity normalised at each time step. (b) $\phi_B$, the HMF angle in the ecliptic plane. Red/blue shaded regions show HMF pointed away/toward the Sun, while the blue/red lines indicate the ideal Parker spiral directions for away/toward polarity HMF. (c) $\theta_B$, the angle of the HMF out of the ecliptic plane. (d) $B_R$ the radial HMF component. (e) $|B|$, the HMF intensity. (f) $V$, the solar wind speed. (g) $n_P$, the proton density. Green vertical bars show intervals of inverted HMF and yellow show intervals of undetermined HMF topology. (1-hour means are shown.)}
    \label{fig:insitu}
\end{figure}

Figure~\ref{fig:insitu} shows 1-hour means of ACE solar wind magnetic field and plasma data. For determining the topology of the HMF over CR2101 we used 64-second suprathermal electron and magnetic field data, following the methodology of \citet{owens17}. Accounting for inverted HMF and single-point sampling uncertainty we derive for the heliospheric open solar flux $\Phi_{SS}$ in the range $449$ to $559  \times 10^{20}$\,Mx, with a most probable value of $482 \times 10^{20}$\,Mx.  Another way of roughly estimating the heliospheric open flux is to use simple averages (an hourly average and a daily average) of the in-situ $B_R$, and then obtain the average of these over the entire rotation.  For the hourly averaged data, we obtain an estimate of $626  \times 10^{20}$\,Mx.  This is almost certainly an overestimate of the open flux, for the reasons described in section \ref{sec_derive_helio}.  For the daily average, we obtain $464  \times 10^{20}$\,Mx. This likely underestimates the open flux, because for these longer-time averages, $B_R$ is canceled in the vicinity of the heliospheric current sheet as well as in regions with folded flux.  These two estimates bracket our most probable value ($482 \times 10^{20}$\,Mx) obtained from the more detailed analysis.


\begin{figure}
    \centering
     \includegraphics[width=\textwidth]{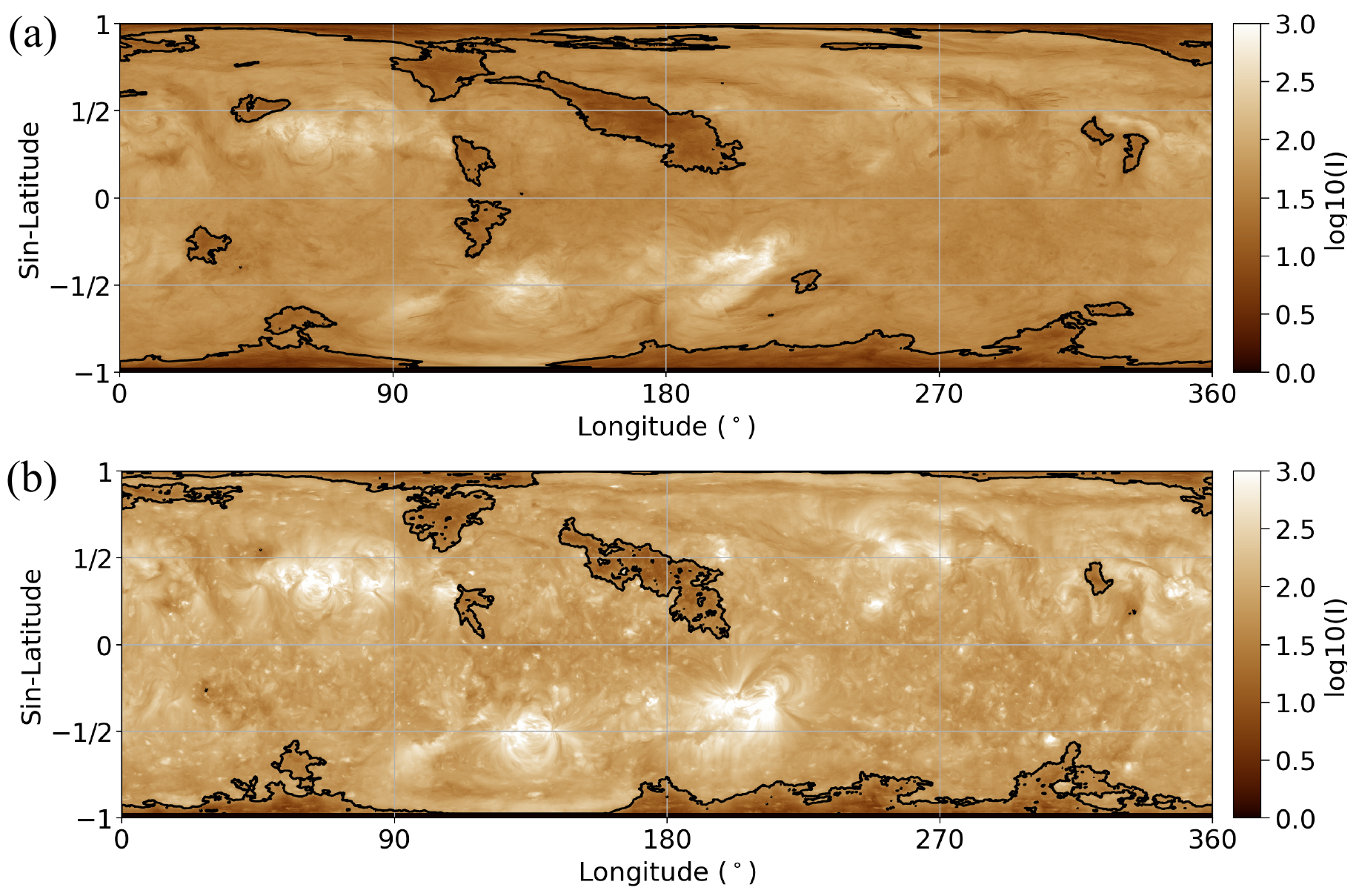}
    \caption{(a) Full-Sun EUV AIA 193\AA\ map and CH detections using PSI-MIDM.  (b) The same as (a) using PSI-SYNOPTIC. }
    \label{fig_psimidm}
\end{figure}

Comparison of the open flux in CHs at the Sun with interplanetary measurements requires CH detection for the entire solar surface. The PSI detection techniques, PSI-SYNCH/SYNOPTIC and PSI-MIDM (see section~\ref{sec_detect_method}), are designed to produce a full-Sun map of EUV and extracted CHs. PSI-SYNCH cannot be used in this capacity for CR2101, because at that time the combined view of STEREO-EUVI and SDO-AIA did not extend over the entire Sun.  PSI-MIDM uses multiple views from AIA, and provides a CH map of the entire Sun over CR2101 shown in Figure~\ref{fig_psimidm}(a). We also employ PSI-SYNOPTIC, which is similar to PSI-SYNCH, but is built up over a solar rotation in order to obtain a full-Sun view (Figure~\ref{fig_psimidm}(b)). (We note that a portion of the south polar region of the Sun was not visible during this period, and we assume that this is open. This lack of visibility likely has a small impact on the estimate of open flux, as described in the Appendix.)

To estimate magnetic flux from the global Sun CH maps, we employ three synoptic maps, namely from HMI, MDI, and GONG, as they are built up over the course of a solar rotation (ADAPT maps are not appropriate as they provide a synchronic representation). Using the two full-Sun maps, PSI-MIDM and PSI-SYNOPTIC, we sum the area of all the detected open field regions, and we sum the fluxes in each CH individually, then take the absolute value of each before summing them together. For PSI-MIDM, we obtain an estimate of $84.0 \times 10^{10}$\,km$^{2}$ for the open field area and $190\times 10^{20}$\,Mx for the open flux, using the HMI synoptic map.  With the MDI synoptic map, the open flux is $216\times 10^{20}$\,Mx, and $160\times 10^{20}$\,Mx for GONG.  All of these values are significantly less than the inferred interplanetary flux of $482\times 10^{20}$\,Mx. The open flux values are even smaller for PSI-SYNOPTIC.  With a detected area of $67.5 \times 10^{10}$\,km$^{2}$, the open fluxes are  $154\times 10^{20}$\,Mx (HMI), $174\times 10^{20}$\,Mx (MDI), and $122\times 10^{20}$Mx (GONG).

When compared to all the other detection methods and input emission data, PSI-MIDM provides the largest area and flux estimates for our selected CH. Yet the estimate of the solar open flux from all detected CHs in CR2101 with PSI-MIDM is well below the interplanetary flux estimate (by a factor of $\approx 2.2$) in the best-case scenario. Our results imply that detected CHs over the entire Sun, regardless of the detection method, contain significantly less flux than is implied by interplanetary observations.  However, there may be greater uncertainty in the detection of CHs in other solar regions, particularly at the Sun's poles.  Another way to test CH detection techniques is to apply them to a model where the true answer is known.  We describe this approach in the following section.

\section{Detection applied to Simulated Emission}
\label{sec_detection_sim_emission}

In section \ref{sec_detect_results}, we found that the uncertainty estimates for CH area and open flux for our well-observed CH are far below the amount required to account for the large difference between our coronal and heliospheric open flux estimates for the entire Sun during this time period. However, there is no ``ground truth'' measurement for the open flux in a CH.  There could be systematic errors related to the geometry and properties of CHs for all detection methods.  From observational data alone, it is difficult to assess how (i) obscuration by nearby overlying loops, (ii) foreshortening of the emission away from disk center, especially near the poles, and (iii) variation in emission intensity over the disk, affect detection of open magnetic flux.  Thermodynamic MHD models \citep[e.g.,][]{lionelloetal2009,downsetal2010,downsetal2013,vanderholstetal2014,mikicetal2018,revilleetal2020} incorporate a realistic energy equation that accounts for anisotropic thermal conduction, optically thin radiative losses, and coronal heating, allowing the plasma density and temperature to be computed with sufficient accuracy to simulate EUV and soft X-ray emission observed from space.  To assess how well detection techniques perform when the true answer is known, we developed an MHD simulation of CR2101 using the Magnetohydrodynamic Algorithm outside a Sphere (MAS) code.  The solution parameters closely resemble those used for the coronal prediction for the August 21, 2017 total solar eclipse \citep{mikicetal2018}; a brief description of these parameters and the computation of simulated emission is described in the Appendix.  We describe the features of the simulation relevant to our CH detection tests in the following section.

\subsection{Properties of the Simulated Corona} 
\label{sec_sim_corona}
\begin{figure}
    \centering
     \includegraphics[width=\textwidth]{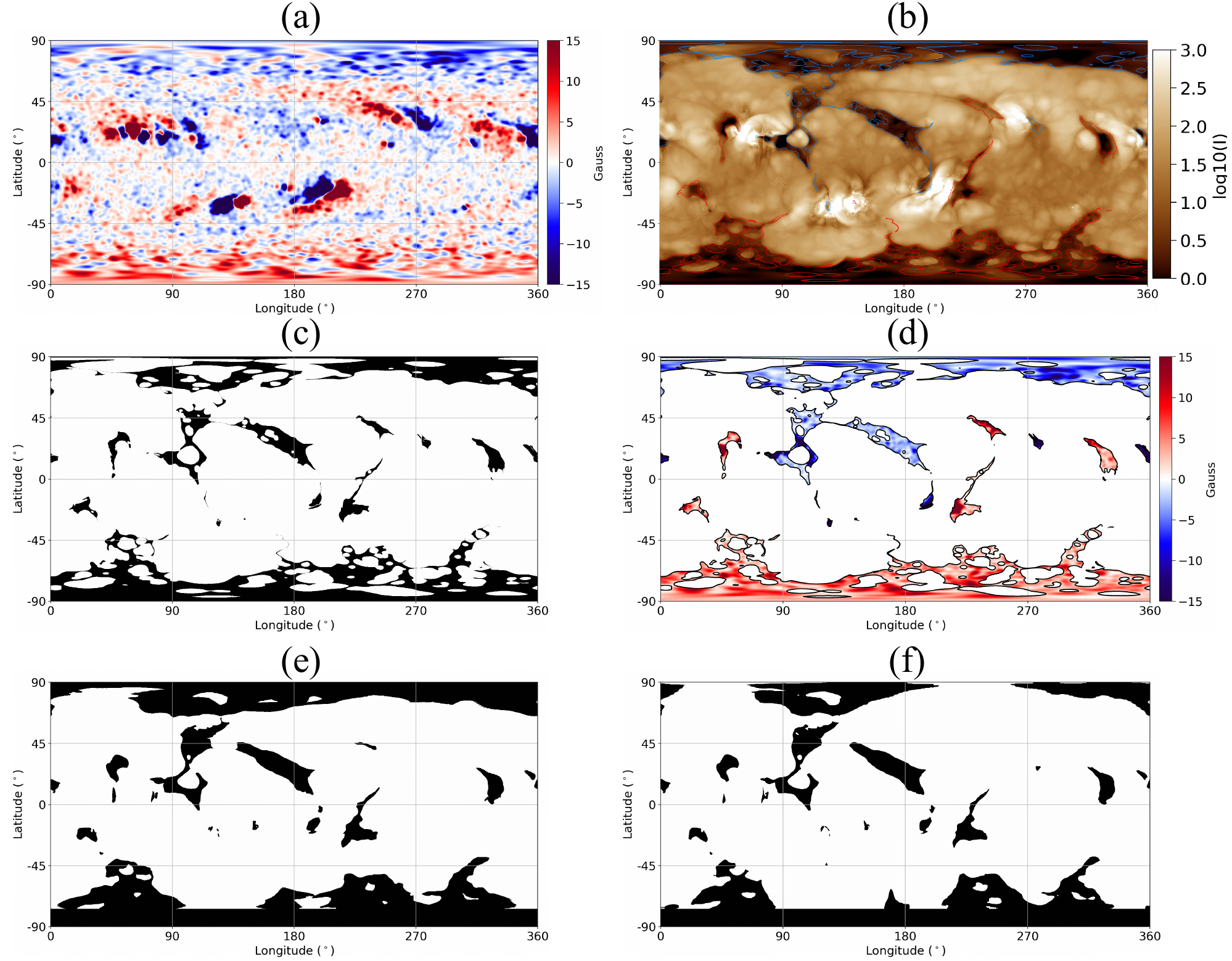}
    \caption{Results from a thermodynamic MHD simulation of the solar corona for CR2101.(a) Map of $B_r$ at the solar surface from an HMI synoptic map, the boundary condition for the calculation. (b) Map of simulated AIA 193\AA\ emission from the model (see text).  The open field regions of the model, calculated by tracing field lines, are overlaid on the EUV map as colored contours.  Red colors indicate positive (outward) polarity, and blue colors indicate negative (inward) polarity. (c) Map of the  open (black) closed (white) field regions of the model, (d) Map of $B_r$ overlaid on the open field map.  This is the open magnetic flux in the model. (e) CHs in the model (black) detected by PSI-MIDM. (f) CHs in the model detected by PSI-SYNOPTIC.}
    \label{fig_thermo}
\end{figure}

Figure \ref{fig_thermo} shows some global diagnostics from the model results.  Figure \ref{fig_thermo}(a) shows the surface magnetic map ($B_r$) used as the boundary condition.  Red indicates positive (outward) polarity ($+B_r$) and blue indicates negative (inward) polarity ($-B_r$).  Figure \ref{fig_thermo}(b) shows a map of the simulated AIA 193\AA\ emission from the model, with contours of the open field regions (red for $+B_r$, blue for $-B_r$) overlaid on the map.  A map of these open/closed field regions is provided in Figure \ref{fig_thermo}(c).   These are the ``target'' areas for our detection methods.  Figure \ref{fig_thermo}(d) shows $B_r$ overlaid on the open field map.  This is the ``true'' open magnetic flux in the model, the target that our detection methods seek to extract.  Figure \ref{fig_thermo}(b) shows that in the model,  in addition to open field regions associated  with dark emission and more unipolar magnetic fluxes, there are also dark regions and open flux next to active regions.  This is in contrast to the observations (Figure \ref{fig_psimidm}), where this dark emission is less apparent near the active regions, but may be obscured by bright active region loops.  Figures \ref{fig_thermo}(e) and (f) show full-Sun CH detections and are described in section \ref{sec_test_detect}.

To create simulated EUV images, we convolve the plasma density and temperature from the model with the SDO/AIA instrument response functions. Synthetic images are created by integrating the 3D volumetric emissivity along the line of sight from a given viewpoint (see the Appendix for more details).  The EUV map in Figure \ref{fig_thermo}(b) was constructed by integrating along radial lines of sight;  \citet{linkeretal2017} described a detection test with PSI-SYNOPTIC on a similar EUV map.  While this comparison yielded useful insights, in general, such a map is more favorable for detection than real images from spacecraft instruments.  

\begin{figure}
    \centering
     \includegraphics[width=\textwidth]{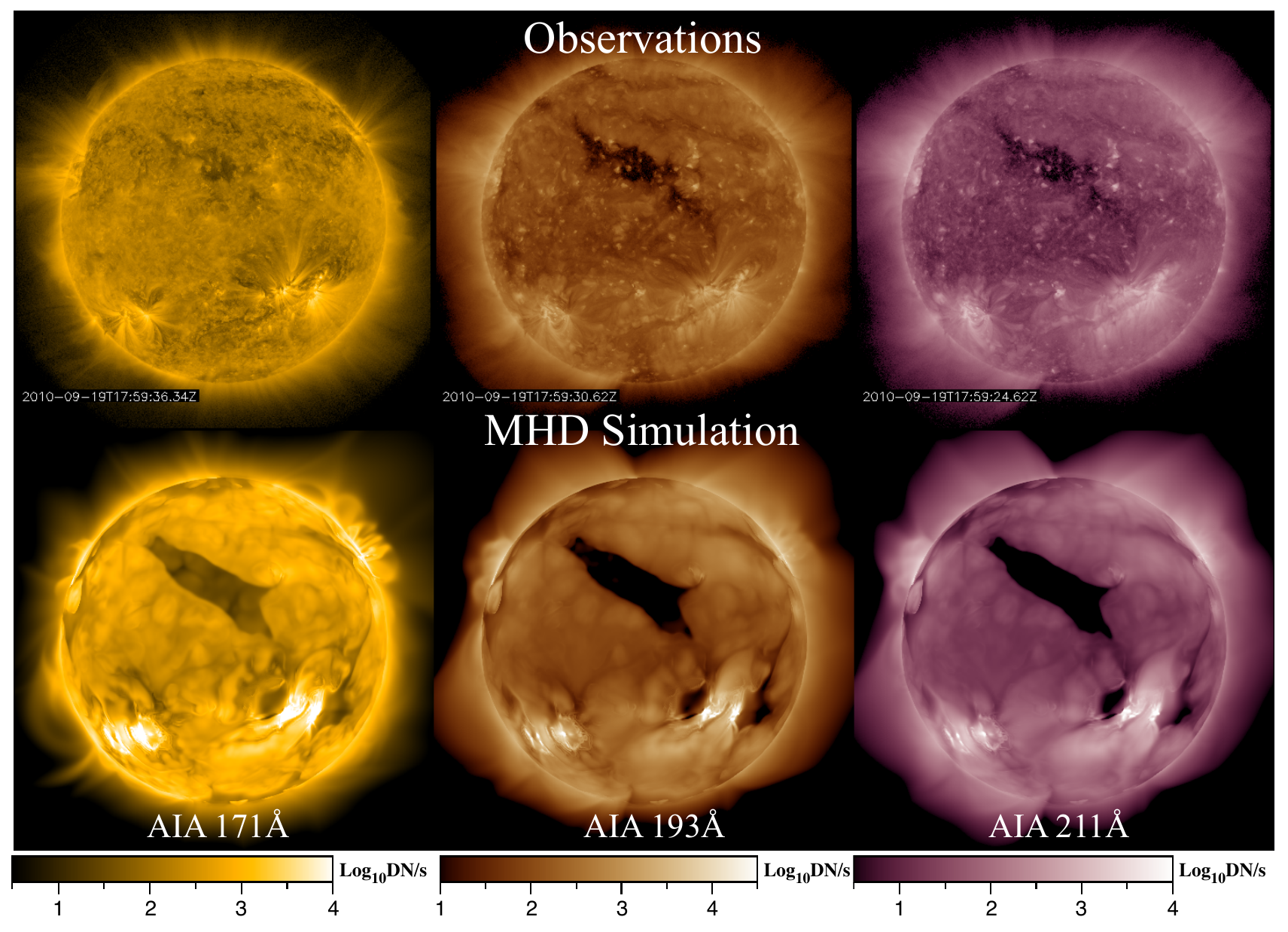}
    \caption{Comparison of observed and simulation AIA emission images on September 19, 2010, at 17:59~UT. Top row: observed AIA emission in 171\AA\ (left), 193\AA\ (middle), and 211\AA\ (right). Bottom row: Simulated emission in the same wavelengths.}
    \label{fig_AIA_sim}
\end{figure}

To provide more realistic conditions to test CH detection techniques, we created a sequence of synthetic emission images in multiple wavelengths from the MHD simulation as observed from the vantage point of SDO/AIA.  Figure \ref{fig_AIA_sim} shows an observational comparison for the date and time of our selected CH in the different SDO/AIA filters. The comparison shows that the model has roughly captured all prominent features of the corona at this time, including the approximate location and size of active regions and CHs. However, the simulated CHs are generally darker and more uniform than in real observations. This is in part caused by the smoothness of the boundary map (Figure \ref{fig_thermo}(a)), which does not include the mixture of small scale parasitic polarities that are prevalent at high resolution (compare with Figure \ref{fig:mag}).  These small scale structures likely contribute to the bright points of emission that tend to ``break up'' CHs.  This effect is likely to be especially prominent in the 171\AA\ line, where small-scale heating processes may dominate the lower temperature plasma and exhibit more structure at smaller scale heights.  The simulated active regions also tend to be less structured than the observed ones.  Conversely, some of the simulated active region emission is over-bright compared to the observations, and this may lead to more obscuration than in observed structures.  These less realistic attributes of the simulated corona are related to resolution/computational cost and properties of the coronal heating model. Further details are provided in the Appendix.  


\subsection{Testing Detection Methods }
\label{sec_test_detect}
\begin{figure}
    \centering
    \includegraphics[width=\textwidth]{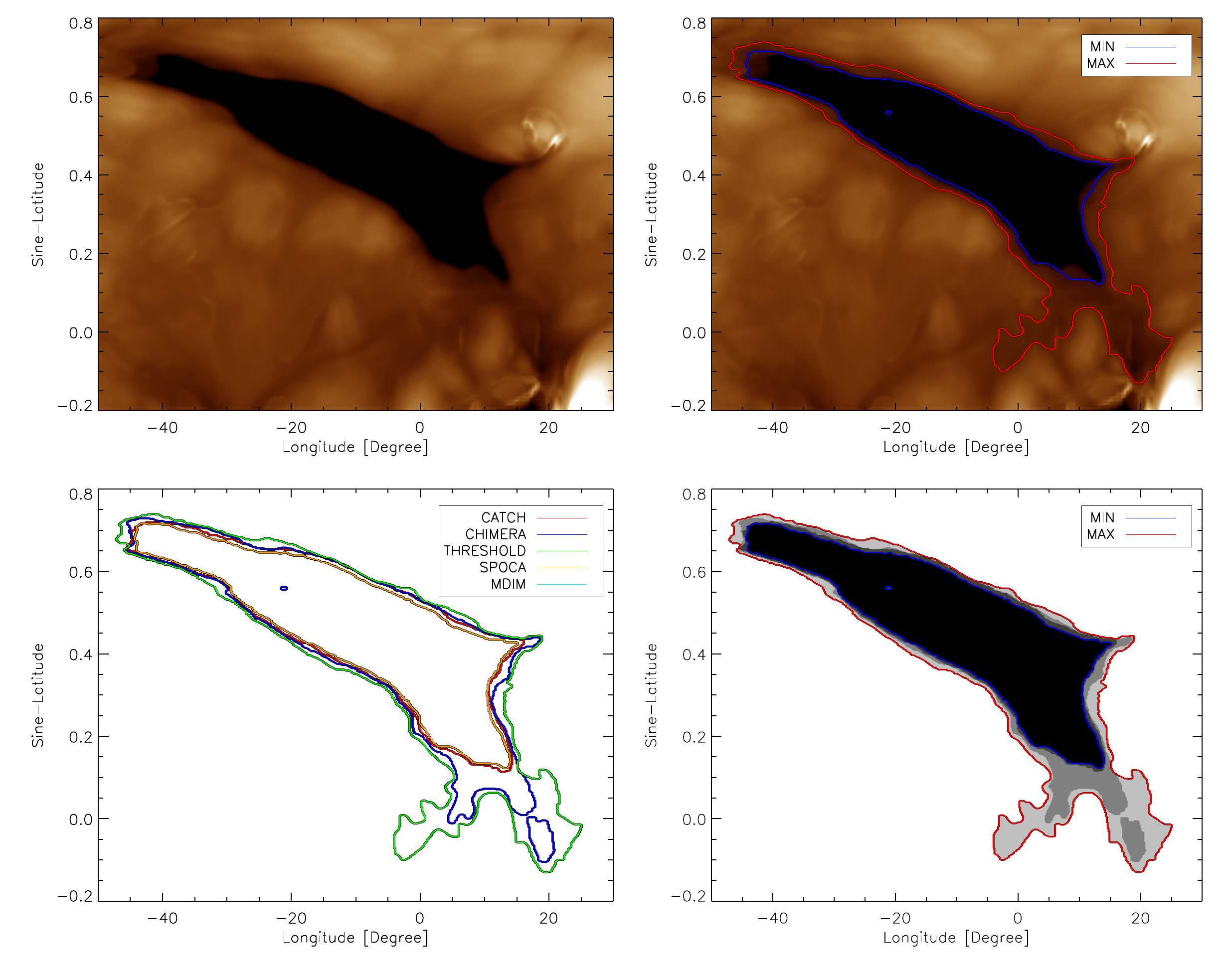}
    \caption{CH boundaries extracted using different methods applied to synthetic AIA/SDO emission, derived from the MHD model. (a) The synthetic AIA/SDO $193$\AA\ filtergram, with the contours of the true open field area shown in cyan.  (b) The same filtergram as (a), overlaid with true open field (cyan), the minimum (dark blue) and maximum (red) boundary constructed by overlaying the 5 individual boundaries.
    (c)  The different boundaries extracted by the means of CATCH (from AIA/SDO 193\AA), CHIMERA (from AIA/SDO 171/193/211\AA), thresholding (with $35\%$ of the median solar disk intensity), SPoCA (AIA 193\AA ), and MIDM (AIA 193\AA\ multiple images), along with the true open field. (d) a stacked map of the 5 binary extractions (where darker pixels represent an agreement of more extractions).}
    \label{fig:synth}
\end{figure}

We applied our CH detection methods to the synthetic AIA images which we created from the simulation, each method using its own processing methodology as if this was a real observation.  The results are shown in Figure \ref{fig:synth}.  The simulated emission in the selected CH (top left) shows less structure than the real CH.  The contours of the true open field regions reveal that magnetic structure in the simulation is more complex.   A closed field region is present within the main body of the CH, but this feature is not revealed in emission, possibly because of issues described in the Appendix.  Contours of the different detection schemes (Figure \ref{fig:synth}, bottom left) all incorporate this region as open.  The CATCH, SPoCA, and PSI-MIDM extracted boundaries are similar, with CHIMERA and THR providing somewhat larger boundaries.  The minimal and maximal boundaries, constructed by overlaying the 5 individual boundaries, along with the true open field boundaries, are shown in the top right of  Figure \ref{fig:synth}.  A stacked map of the detections is shown in the bottom right.  To estimate the magnetic fluxes in the detections, we assume the values in the boundary map that was used in the simulation (Figure \ref{fig_thermo}(a)), i.e., we assume there is no error in magnetic flux incorporated in the detections.

The numerical results for the detections, along with the true values, are provided in Table \ref{tab:model_ch}.  The average detected area of the CH ($11.5 \times 10^{10}$\,km$^2$) is considerably larger than the true area ($7.4 \times 10^{10}$\,km$^2$, but the average strength of $B_r$ in the detected areas is  smaller in absolute value ($-2.1$\,G) than the true value ($-3.1$\,G). This occurs because regions that were miss-identified as open by the detection methods had primarily mixed polarity magnetic flux, lowering the magnitude of the average value.  The resulting average estimate for the open magnetic flux from this CH is $24.2 \times 10^{20}$\,Mx for all the detections, only about 7\% above the true value.
All but one of the methods overestimated the open magnetic flux in the CH. The reasons for this are discussed further in section \ref{sec_overestimate}, but probably occur because this is a relatively simple CH, situated away from any strong active regions.  We will see this is generally not the case when detecting open magnetic flux from all CHs (i.e.\,over the full Sun).
\begin{table}
\centering
\caption{Extracted CH parameters: area ($A$), intensity in 193\AA~image data, (I$_{\rm 193}$), location of the center of mass, (CoM, longitude and latitude), magnetic field strength ($B$), and magnetic flux ($\Phi$), from the synthetic 193\,\AA\ filtergrams using CATCH, CHIMERA, THR, SPoCA and PSI-MIDM. The magnetic field properties were derived using the input magnetic map for the MHD simulation. }
\label{tab:model_ch}
\centering
\begin{tabular}{l | c c c c c | c || c } 
 & CATCH & CHIMERA & THR & SPoCA & PSI-MIDM & Mean & True \\ \hline
A [$10^{10}$ km$^{2}$]& $10.5$& $12.4$& $15.5$ & $9.5$ & $9.7$ &$11.5\pm2.5$& $7.4$  \\
I$_{193}$ [DN] & $62.4$& $66.2$& $69.0$ & $60.0$ & $60.2$ &$63.6\pm4.0$& $61.2$  \\
CoM$_{\mathrm{\textsc{lon}}}$ [$^{\circ}$] & $-10.4$& $-8.1$& $-6.4$ & $-10.3$ & $-10.5$ &$-9.2\pm1.8$& $-9.8$  \\
CoM$_{\mathrm{\textsc{lat}}}$ [$^{\circ}$] & $26.6$& $23.9$& $21.2$ & $26.4$ & $26.5$ &$24.9\pm2.4$& $26.2$  \\ 
B [G]& $-2.3$& $-2.0$& $-1.7$ & $-2.3$ & $-2.4$ &$-2.1\pm0.3$& $-3.1$  \\
$\Phi$ [$10^{20}$ Mx]& $-24.0$& $-25.3$& $-26.6$ & $-21.6$ & $-23.2$ &$-24.2\pm1.9$& $-22.6$  \\ \hline

\hline
\end{tabular}
\end{table}

To investigate how well CH detection methods can estimate the total open magnetic flux, we again employ the PSI-MIDM and PSI-SYNOPTIC  algorithms, as we did for the observed CHs of CR2101.  Figures \ref{fig_thermo}(e) and (f) show the detected CHs using these two methods.  Comparison with Figures \ref{fig_thermo}(c) and (d) shows that  the algorithms generally detect most of the true open field regions.  The CH structure identified by the methods is generally simplified compared to the true open fields, and areas are overestimated in a number of regions, especially the poles.  Some smaller CHs are missed or not fully captured  and as these areas are associated with active regions, they contain a disproportionate amount of 
open flux.  

For the PSI-MIDM method, we find that the total Sun detected CH area is $83.91 \times 10^{10}$\,km$^2$ and the detected open flux is $285 \times 10^{20}$\,Mx.  The PSI-SYNOPTIC method finds a detected CH area of $67.98 \times 10^{10}$\,km$^2$ and detected open flux of $240 \times 10^{20}$\,Mx.  The true open field area was $62.71 \times 10^{10}$\,km$^2$ and true open flux was $387 \times 10^{20}$Mx.  Both methods overestimate the total areas of all CHs ($33.8\%$ by PSI-MIDM and $8.4\%$ by PSI-SYNOPTIC) but underestimate the total open flux (PSI-MIDM captures $73.7\%$ of the flux and PSI-SYNOPTIC captures $62.0\%$).  These underestimates are discussed further in the following section.

\section{Discussion}
\label{sec_discussion}

\subsection{CH Detection on a Well-Observed CH}
Our comparison of detection methods for the area of the observed CH reveals a standard deviation of $\sigma_{\rm A}\approx$~26\% from the mean value, which we estimate to be the approximate uncertainty in the methods. The variability in the magnetic fluxes from different magnetic field maps raises this uncertainty to $\sigma_{\rm A}\approx~43-46\%$. In the methods comparison performed on the simulated CH, the standard deviation for the detected areas was similar to the observed case, about $21\%$.  However, the mean of these values ($11.5\times 10^{10}$\,km$^2$) was actually $36\%$ greater than the true value ($7.4\times 10^{10}$\,km$^2$).  All but one of the methods overestimated the open flux in the simulated CH, but the standard deviation in the open flux was much smaller ($8.4\%$) than for areas. The actual error of the mean open flux compared with the true value was even less ($7\%$). This reflects the fact that if mixed polarity regions are mis-identified as open, they don't contribute as much to the flux estimate because the opposite polarities cancel in the integration.  

\subsection{Detection of the Global Open Flux}
Tables \ref{tab:Observed_full_sun} and \ref{tab:Model_full_sun} summarize our full-Sun detections for both observations and the model. 
The open flux inferred from heliospheric observations for this time period was in the range $449-559 \times 10^{20}$\,Mx, with a most probable value of $482 \times 10^{20}$\,Mx, corresponding to $B_{\rm R} = 1.71$\,nT at 1 AU.  The full-Sun detections, regardless of the magnetic map used, greatly underestimate these values. The highest estimate ($0.77$\,nT) comes from PSI-MIDM with the MDI synoptic magnetic field map, and this detection was actually an outlier for the detections on the selected individual CH. The uncertainties we estimated for the area and open flux of the selected well-defined CH are likely less than for full-Sun detections, where factors such as obscuration and viewing geometry play a larger role. Our full-Sun detections on the model corona at least partially account for these aspects, and indeed show that the methods, while overestimating the CH area, actually underestimate the global open flux (e.g., the true open flux is $35\%$ greater than the PSI-MIDM estimate). However, even applying this factor to the PSI-MIDM full-Sun detection of the observations leaves us well short of the estimated heliospheric interplanetary flux.  
\begin{table}
\centering
\caption{Areas and open fluxes derived from full-Sun detections of observed CHs for CR2101.}
\label{tab:Observed_full_sun}

\centering
\begin{tabular}{l c | c c | c } 
  Method & Magnetic Map &  Area [$10^{10}$ km$^{2}$] & Open Flux [$10^{20}$ Mx]  & Heliospheric [$10^{20}$ Mx]  \\
  & & & ($B_r$ at 1 AU) & ($B_r$ at 1 AU) \\ \hline
\multirow{3}{*}{MIDM}  & HMI& \multirow{3}{*}{$84.0$}& $191$  ($0.68$\,nT)  & \multirow{6}{*}{$482$  ($1.71$\,nT)}  \\
 & MDI && $216$  ($0.77$\,nT) &    \\
 & GONG && $160$  ($0.57$\,nT) &    \\ \cline{1-4}
\multirow{3}{*}{SYNOPTIC} & HMI& \multirow{3}{*}{$67.5$}& $154$  ($0.55$\,nT) &    \\ 
 & MDI  && $174$  ($0.62$\,nT) &    \\
 & GONG  && $122$  ($0.43$\,nT) &    \\
\hline
\end{tabular}
\end{table}

\begin{table}
\centering
\caption{Areas and open fluxes derived from full Sun detections of simulated CHs for CR2101.}
\label{tab:Model_full_sun}

\centering
\begin{tabular}{l c | c c | c  c } 
 & Method &  Area [$10^{10}$ km$^{2}$] & Open Flux [$10^{20}$ Mx] ($B_r$ at 1 AU)& True Area &True Flux \\ \hline
& MIDM & $83.9$& $284$  ($1.01$ nT) & \multirow{2}{*}{$62.7$}&  \multirow{2}{*}{$387$ ($1.37$\,nT)} \\
& SYNOPTIC & $68.0$& $240$  ($0.85$ nT) && \\
\hline
\end{tabular}
\end{table}

\subsection{Overestimates of Open Flux in Individual CHs}
\label{sec_overestimate}
There are two primary sources to the overestimate of the area and open flux on the individual, simulated CH.  The first can be seen by comparing the true open field contour (cyan) in Figure \ref{fig:synth} with the simulated emission and all of the extracted CH boundaries in the figure.  Pockets of closed field appear dark in emission in the figure, and are indistinguishable from open field to the detection methods.  This may be less likely to occur on the real Sun, where these small-scale loops may appear brighter in emission than they do in the model.  The absence of emission here may be due to deficiencies in the coronal heating model for small-scale loops (see Appendix).

The second source of overestimation occurs because, as implemented here, the methods do not account for the coronal height at which the bulk of the emission begins to form in the EUV lines used in the detection \citep[estimated to be about $1.01$\,R$_{\sun}$ for 193\AA\ and 195\AA\ emission,][see section 4.2 and Figure 18]{caplanetal2016}.  In general, the magnetic field expands with height and the CH has a larger area at the height of detection than at its magnetic source in the photosphere.  Simply projecting the CH area downward on the photosphere captures a larger area than the actual magnetic source.  In the model, the plasma at chromospheric temperature is artificially thick and the 193\AA\ emission forms at $1.02$\,R$_{\sun}$.  Calculating the area of the true open field at this height, we find that this rises to $9.3 \times 10^{10}$\,km$^2$, much closer to the detected areas (especially for SPoCA and PSI-MIDM).  This result suggests that CH detection methods may be able to improve the estimates of the open magnetic flux by using a potential field model to extrapolate $B_r$ to the height at which the emission forms, slightly lowering the flux estimate.

\hfil\break
\subsection{Underestimates of Total Open Flux}
The full-Sun detections, when applied to the model corona, underestimate the total open flux, with PSI-MIDM accounting for $73.7\%$ of the flux and PSI-SYNOPTIC accounting for $62.0\%$.  The reason for these underestimates can be seen by comparing the true open field regions (Figures \ref{fig_thermo}(c) and (d)) with the detections (Figures \ref{fig_thermo}(e) and (f)).  There are several smaller-scale open field regions that are under-detected or completely missed in the extracted CH boundaries.  These often are in the proximity of active regions, which contain significant amounts of magnetic flux.  For example, the two open field regions near $270^\circ$ longitude and $45^\circ$ latitude are almost completely missed in the extracted CH boundaries but account for $6\%$ of the total open flux in the model.   \citet{linkeretal2017} and  \citet{caplanetal2019} found that open flux was underestimated by these detection methods for the same reasons.  

\subsection{Implications for the Open Flux Problem}
Our comparisons of detection methods on our selected CH, for both the observed and simulated cases, show reasonable agreement between the methods.  Uncertainty from the different magnetic map products contributes as much to the variability as the detections themselves.  The average error in the detected open flux for the simulated CH was relatively small ($7\%$).  The full-Sun detections for the observed case all greatly underestimated the open flux deduced from in-situ measurements, with the largest estimate still a factor of 2.2 smaller than the interplanetary value.   The schemes also underestimated the open flux for the model corona.  However, the model's true open flux was only $35\%$ greater than the estimate from the PSI-MIDM detection.  

The under-detection of open flux within the traditionally described CH areas may well contribute to the open flux problem, however, it seems unlikely that it is the only reason and, therefore, cannot resolve it. 

We note that the open flux in the simulated corona was relatively close ($\approx 80\%$) to the in-situ value. However, the open field regions in the model at mid-latitudes and near active regions are larger and more obvious than in the observations.  If regions like this exist on the real Sun and contribute to the open flux, they would have to be considerably more obscured than occurs in the model.

One possible resolution to the open flux problem is that the under-detection of open flux (e.g., near active regions), in combination with systematic underestimates of magnetic flux by magnetographs (either everywhere on the Sun, or just at the poles), could account for the missing open flux.  In this regard, the behavior of CHs at the poles could be especially important, and our present observational views of the Sun's poles limit our ability to resolve this question. The latter part of the Solar Orbiter mission, when the spacecraft will reach latitudes of $\sim$30$^\circ$, may yield clues to the importance of the polar contribution.  Ultimately, a mission that fully images the Sun's poles \citep[such as the Solaris mission,][]{hassleretal2019,hassleretal2020} can resolve the contribution of the Sun's polar regions to the open flux.

A second possibility is that a significant portion of the open flux is rooted at the Sun, but continually undergoes interchange reconnection, and the mixture of open and closed field lines are not obviously dark in emission.   
While interchange reconnection has been advocated as an explanation for the origin of the slow solar wind \citep[e.g.,][and references therein]{abboetal2016}, it is not clear what emission properties the plasma on these field lines would possess.  Therefore, with the present state of the theories/models, it is difficult  to either completely confirm or falsify this idea from observations alone.  An advanced model that simulates the time-dependent evolution of the corona and demonstrates the observed emission properties would seem to be necessary to progress beyond the present qualitative arguments.  

A third possibility is that the disparity between observed coronal and heliospheric open flux is not related to solar observations, but to the behavior of the interplanetary magnetic field.  The discovery of long intervals of ``switchbacks'' in the interplanetary magnetic field from PSP \citep{baleetal2019,kasperetal2019} suggests that folded flux could be more ubiquitous than previously thought, and lead to increases in the magnitude of $B_{\rm R}$ measured in-situ at increasing distance (i.e., 1 AU) from the Sun \citep{macneil20}.  However, comparison of PFSS and MHD models with PSP observations \citep{badmanetal2020,rileyetal2021} suggest that the models significantly underestimate the field strength even at the perihelion distances that PSP has reached thus far, though more detailed accounting for switchbacks may be necessary.  Large amounts of disconnected flux in the heliosphere could also account for the missing open flux.  This has generally been considered unlikely \citep{crooker_pagel2008}, but recent PSP observations of reconnection in the heliospheric current sheet \citep{lavraudetal2020,phanetal2020} indicate that this process appears to be more prevalent than previously thought.


\section{Summary}
\label{sec_summary}
We have investigated CH detection techniques to characterize the uncertainty in characterizing CH area and open flux from observational EUV data.  Starting from a well-observed, near-disk center CH, we applied six different detection methods to deduce the area and open flux.  We also applied a single method to five different EUV filtergrams for this CH.
Open flux was calculated for all of the detections using five different magnetic maps.  Using the standard deviation as a measure of the uncertainty, we find that the uncertainty in the estimate of open flux for this particular CH was $\approx 26\%$. When including the variability in the different magnetic data sources, this uncertainty rises to $43-46\%$.  We used two of the methods to characterize the area and open flux for all CHs during CR2101. We find that the open flux is greatly underestimated compared to the value inferred from in-situ measurements, by a factor of 2.2--4. As there is no ``ground truth'' measurement of open flux in CHs, we tested our detection techniques on simulated emission images from a thermodynamic MHD model of the solar corona for this time period, where the true values of the model are known. The variability in the detected area of the simulated CH is similar to the observed case. The methods generally overestimate the area and open flux in the simulated CH, but the average error in the flux is only about $7\%$. The full-Sun detections on the simulated corona underestimate the model open flux (by $35\%$ for PSI-MIDM), but this factor is well below what is needed to account for the missing flux in the observations.  Our results imply that under-detection of open flux in what are traditionally considered to be coronal holes likely contributes to the open flux problem, but is unlikely to resolve it.


\acknowledgments 
We thank the International Space Science Institute (ISSI, Bern) for the generous support of the ISSI team “Magnetic open flux and solar wind structuring in interplanetary space” (2019-2021). JAL, RMC, and CD were supported by the NASA Heliophysics Guest Investigator program (grant NNX17AB78G), the NASA HSR program (grant 80NSSC18K0101 \& 80NSSC18K1129), AFOSR (contract \# FA9550-15-C-0001), the NSF PREEVENTS program (grant ICER1854790)), and the STEREO SECCHI contract to NRL (under subcontract N00173-19-C-2003 to PSI).  Computational resources were provided by NASA's NAS (Pleiades) and NSF's XSEDE (TACC \& SDSC).
BV acknowledges the financial support by the Croatian Science Foundation under the project 7549 ``Millimeter and submillimeter observations of the solar chromosphere
with ALMA.'' SJH was supported, in part, by the NASA Heliophysics Living With a Star Science Program under Grant No. 80NSSC20K0183. EA acknowledges support from the Academy of Finland (Postdoctoral Researcher Grant 322455). MO is funded by part-funded by Science and Technology Facilities Council (STFC) grant numbers ST/R000921/1 and ST/V000497/1. CS acknowledges support from the Research Foundation -- Flanders (FWO, strategic base PhD fellowship No. 1S42817N), and from the NASA Living With a Star Jack Eddy Postdoctoral Fellowship Program, administered by UCAR's Cooperative Programs for the Advancement of Earth System Science (CPAESS) under award no. NNX16AK22G. ES acknowledges support from a PhD grant awarded by the Royal Observatory of Belgium. JP acknowledges support from the SolMAG project (ERC-COG 724391) funded by the European Research Council (ERC) in the framework of the Horizon 2020 Research and Innovation Programme, and the Finnish Centre of Excellence in Research of Sustainable Space (Academy of Finland grant number 312390). ICJ acknowledges support from a PhD grant awarded by the Royal Observatory of Belgium.

\appendix

\section{Thermodynamic MHD model}
\label{sec_sim_emission}


We developed an MHD simulation of CR2101 using the Magnetohydrodynamic Algorithm outside a Sphere (MAS) code.  The method of solution, including the boundary conditions, has been described previously \citep[e.g.,][]{mikiclinker94,linkeretal1999,lionelloetal1998,lionelloetal2009}.  The solution parameters closely resemble those used for the coronal prediction for the August 21, 2017 total solar eclipse \citep{mikicetal2018}, including a Wave-Turbulence-Driven (WTD) model of coronal heating \citep{downsetal2016}.  The simulation utilized a $288 \times 327 \times 699$ nonuniform $r,\theta,\phi$ mesh, with the smallest radial cells of  $\Delta r = 0.00043 R_S$ ($\approx 300$ km) near the solar surface and $\Delta r = 0.83R_S$ at the outer boundary of $30 R_S$.  For the co-latitude mesh, $\Delta \theta = 0.51^\circ$ near the equator and $0.86^\circ$ near the poles, and for the longitudinal mesh, $\Delta \phi =  0.51^\circ$ was uniform.
The boundary condition was derived from the HMI LOS synoptic map for CR2101, available at \blue{\url{http://jsoc.stanford.edu/HMI/LOS_Synoptic_charts.html.}}  As in \cite{mikicetal2018}, we multiplied the HMI map by a factor of 1.4 to account for the difference between HMI and MDI data \citep{liu12}, which was previously used to calibrate the heat flux in the model.  We divide out this 1.4 factor prior to computing the true and estimated magnetic fluxes in the detected CHs described in section \ref{sec_test_detect}, to make these more comparable with the values derived from observations.

The instrument response functions were developed using the AIA v6 calibration, the CHIANTI 8.0.2 database \citep{delzanna15} and the CHIANTI hybrid abundances \citep{fludra99}, based off \citep{schmelzetal2012}.   Synthetic images are created by integrating the 3D volumetric emissivity along the line of sight from a given viewpoint.  The EUV map in Figure \ref{fig_thermo}(b) was constructed by integrating along radial lines of sight;  \citet{linkeretal2017} described a detection test with PSI-SYNOPTIC on a similar EUV map.  While this comparison yielded useful insights, in general, such a map is more favorable for detection than real images from spacecraft instruments.  While obscuration of CHs from bright loops can still occur, this line of sight occurs only at disk center.  Away from disk center (especially at higher latitudes) more obscuration may occur, and polar regions are especially foreshortened.  

To test CH detection techniques under more realistic conditions, with a dataset more akin to those actually produced by AIA, we created a sequence of synthetic emission images in multiple wavelengths from the MHD simulation as observed from the vantage point of SDO/AIA.   The B0 angle (i.e., the heliographic latitude of the central point of the solar disk) of the Sun is included in the geometry.  A set of images was created approximately every six hours, for a total of 111 image sets. As described in section \ref{sec_sim_corona}, smoothing of the magnetic map reduces the presence of small-scale, mixed polarities, and these provide important contributions to the complexity of real emission images.  The amount of smoothing of the map is in turn determined by the available resolution for the MHD simulation, which strongly influences the computational cost.  A second simplification of the model is the attempt to describe all of coronal heating with the simplified WTD description \citep[for details, see][]{downsetal2016}.  The origin of coronal heating is, of course, controversial.  The WTD mechanism, even if proven generally valid, may not be applicable to heating at all coronal scales, as small-scale heating may be dominated by other mechanisms - this may be more important for the 171\AA\ line, that has contributions from lower temperature plasma and exhibits more structure at smaller scale heights.  This effect can only be explored by performing much higher resolution simulations than the one we employ here.  Furthermore, to capture the solar atmosphere's thin transition region while still modeling the vast scales of the solar corona, the simulation artificially broadens the transition region by modifying the thermal conduction and radiative losses at lower temperatures.  This approach \citep{lionelloetal2009,mikicetal2013} has been shown to accurately reproduce coronal solutions at higher temperatures  (for this case, $\ > 400,000$K) but can significantly modify the density (and thus emission) at lower temperatures - this effect is again most likely to influence simulated 171\AA\ emission.  Accuracy at lower temperatures can be provided at the cost of smaller cells in the transition region.

Despite the aforementioned shortcomings, the simulated emission images still provide a robust test of CH detection techniques, including obscuration by overlying structures, orders of magnitude differences in emission intensity between different portions of the solar disk, and realistic geometry.  
As with the observations, a small portion of the southern pole is not visible from our simulated viewpoint.  In our full-Sun detection tests, we assumed this region was open, the same as we did in the observed case.  This turns out to impact the open flux estimate by less than 1\%, compared to the true value of the model.  Therefore, we expect this assumption to also have minimal impact on the full-Sun estimates for the observed case.

\bibliographystyle{apj}
\bibliography{ISSI_bib}

\newpage

\end{document}